\documentclass{emulateapj}
\usepackage{apjfonts}
\usepackage{epsf}

\slugcomment{Accepted in ApJ}

\newcommand\Rein{R_{\rm ein}}

\newcommand\mumin{\mu_{\rm min}}
\newcommand\mumax{\mu_{\rm max}}
\newcommand\Mdwarf{M_{\rm dwarf}}
\newcommand\Mclus{M_{\rm clus}}
\newcommand\Sclus{\sigma_{\rm clus}}

\newcommand\uu{{\bf u}}
\newcommand\xx{{\bf x}}

\newcommand\Dt{\Delta\theta}

\newcommand\reffig[1]{Figure~\ref{fig:#1}}

\newcommand\reftab[1]{Table~\ref{tab:#1}}
\newcommand\refsec[1]{\S~\ref{sec:#1}}

\begin{document}

\title{Gravitational lensing magnification without multiple imaging}

\author{
  Charles R.\ Keeton\altaffilmark{1,2,3},
  Michael Kuhlen\altaffilmark{4},
  \& Zolt\'an Haiman\altaffilmark{5}
}

\altaffiltext{1}{Astronomy \& Astrophysics Department,
University of Chicago, 5640 S.\ Ellis Ave., Chicago, IL 60637}
\altaffiltext{2}{Hubble Fellow}
\altaffiltext{3}{Department of Physics \& Astronomy, Rutgers University,
136 Frelinghuysen Road, Piscataway, NJ 08854}
\altaffiltext{4}{Department of Astronomy \& Astrophysics,
University of California, 1156 High St., Santa Cruz, CA 95064}
\altaffiltext{5}{Department of Astronomy, Columbia University,
550 W.\ 120th St., New York, NY 10027}

\begin{abstract}
Gravitational lensing can amplify the apparent brightness of distant
sources.  Images that are highly magnified are often part of
multiply-imaged systems, but we consider the possibility of having
large magnifications without additional detectable images.  In rare
but non-negligible situations, lensing can produce a singly highly
magnified image; this phenomenon is mainly associated with massive
cluster-scale halos ($\gtrsim\!10^{13.5}\,M_\odot$).  Alternatively,
lensing can produce multiply-imaged systems in which the extra images
are either unresolved or too faint to be detectable.  This phenomenon
is dominated by galaxies and lower-mass halos
($\lesssim\!10^{12}\,M_\odot$), and is very sensitive to the inner
density profile of the halos.  Although we study the general problem,
we customize our calculations to four quasars at redshift $z \approx
6$ in the Sloan Digital Sky Survey (SDSS), for which \citet{richards}
have ruled out the presence of extra images down to an image splitting
of $\Dt=0\farcs3$ and a flux ratio of $f=0.01$.  We predict that
9--29\% of all $z \approx 6$ quasars that are magnified by a factor of
$\mu>10$ would lack detectable extra images, with 5--10\% being true
singly-imaged systems.  The maximum of 29\% is reached only in
the unlikely event that all low-mass ($\lesssim\!10^{10}\,M_\odot$)
halos have highly concentrated (isothermal) profiles.  In more realistic
models where dwarf halos have flatter (NFW) inner profiles, the
maximum probability is $\sim\!10$\%.  We conclude that the probability
that {\em all four} SDSS quasars are magnified by a factor of 10 is
$\lesssim 10^{-4}$.  The only escape from this conclusion is if there
are many (>10) multiply-imaged $z \approx 6$ quasars in the SDSS database
that have not yet been identified, which seems unlikely.  In other words,
lensing cannot explain the brightnesses of the $z \approx 6$ quasars,
and models that invoke lensing to avoid having billion-$M_\odot$ black
holes in the young universe are not viable.
\end{abstract}


\section{Introduction}

Two conspicuous aspects of gravitational lensing are the ability to
(i) produce multiple images, and (ii) magnify the apparent brightness
of a distant background source.  Large magnifications generally require
that the projected density along the line of sight be of order the
critical surface density for lensing, which in turn implies a precise
alignment of the observer, the lens, and the source.  The same condition
generally leads to the production of multiple images, so in most cases
one expects highly magnified objects to have at least one companion
lensed image.

The connection between magnification and multiple imaging can be
important in a variety of contexts.  An important example, which
serves as the main motivation for this paper, is the recent discovery
of bright quasars at redshifts as high as $z \sim 6$ in the Sloan
Digital Sky Survey \citep[SDSS; see][]{fetal00,fetal01,fetal03}.
If these quasars are not lensed or beamed (see \citealt{hc02} and
\citealt*{wmj03} for arguments to justify both assumptions), they
are inferred to be very luminous ($M_B \sim -27$).  Assuming further
that they shine at the Eddington limit of their resident black holes
(BH), these BHs must have masses of several $\times 10^9\,M_\odot$. 

Having such massive BHs at such an early stage in the evolution of the
universe presents a challenge to models where massive BHs grow mainly
by gas accretion that is itself Eddington limited \citep{hl01}.  Even
in the context of hierarchical structure formation models, where
mergers of several BHs can contribute to the build-up of the mass, the
initial seeds are required to be present as early as $z \gtrsim 15$
\citep{hl01}.  The problem is exacerbated if BH--BH mergers result in
the ejection of BHs from the shallow dark matter potential wells at
high redshift because of a large recoil following the emission of
gravitational waves.  In a recent model that includes this effect,
\citet{zh04} found that BHs can grow by mergers and accretion to at
most a few $\times 10^8~{\rm M_\odot}$ by redshift $z=6.4$ without
a super-Eddington phase --- a short-fall by a factor of $\sim\!10$
relative to the BH masses inferred from observations.

If the high-redshift quasars were magnified by gravitational lensing
by a factor of $\mu \gtrsim 10$, this could alleviate the need
for a super-Eddington growth phase to explain such massive early BHs
(since the inferred mass scales as $\mu^{-1}$ under the assumption
that the quasar is shining at the Eddington luminosity).  Although
the lensing optical depth along a random line of sight to $z \sim 6$
is known to be small \citep[$\sim\!10^{-3}$; e.g.,][]{csk98,bl00},
magnification bias can significantly boost the probability of strong
lensing in a real, flux-limited survey.  If the intrinsic (unlensed)
quasar luminosity function at $z \sim 6$ is steep, and/or it extends
to faint magnitudes, the probability of strong lensing for the SDSS
quasars could even be of order unity \citep*{CHS,WL02}.  However, for
a population of isothermal sphere lenses, all magnifications $\mu > 2$
are associated with multiple imaging, and in most cases the angular
separation between the images is more than $0\farcs3$ \citep{CHS}.
Recent HST observations of the highest redshift quasars have shown no
evidence for additional images of any of the $z \approx 6$ sources
down to a splitting angle of $0\farcs3$ \citep{richards}, which
effectively rules out the hypothesis that the quasars are all highly
magnified by {\em isothermal sphere} lenses.

The obvious question is whether the SDSS quasars could be magnified
by lenses with a more complicated (and indeed more realistic)
lens potential, without producing multiple detectable images.
\citet{WL02} showed that microlensing by stars within lens
galaxies can permit magnifications as high as $\mu \sim 10$ for
singly-imaged quasars, but the probability for $\mu > 2$ is still
very low ($<$0.5\% even when magnification bias is included).
Another question is whether departures from spherical symmetry
significantly affect the results.  One goal of this paper is to
study ellipticity in the lens galaxy and tidal shear from objects
along the line of sight, both of which are common in observed
multiply-imaged systems \citep*[e.g.,][]{KKS,witt,holder}.
Ellipticity and shear are known to modify the full magnification
distribution for multiply-imaged sources
\citep[e.g.,][]{BK87,finch,huterer}; but what happens when we
restrict attention to sources without multiple detectable images
is not known.  More generally, our goal is to present a thorough
and general study of the magnifications that can be produced by
lenses with different radial profiles and angular shapes, without
creating multiple detectable images.  We simultaneously consider
both true singly-imaged systems, as well as multiply-imaged
configurations where the images are too close to be resolved or
the extra images are below some reasonable detection threshold.

In this thorough but technical study, let us not lose sight of
the bottom line:  We find that the {\em maximum} probability that a
$z \approx 6$ quasar is magnified by at least a factor of 10 without
having a second detectable image is 29\% (see \reftab{frac}).
Moreover, this maximum is reached only in the unlikely event that
halos down to arbitrarily low mass have highly concentrated (singular
isothermal sphere) profiles.  In more realistic models of the lensing
population, where dwarf halos have flatter (NFW) inner profiles,
the maximum probability is $\sim\!10$\%.

While our analysis is specifically prompted by the SDSS quasars, it
should be applicable to other objects for which significant lensing
amplification would be important, such as high-redshift galaxies
discovered in ``blank'' fields\footnote{Magnification of high-redshift
sources by foreground cluster lenses is certainly interesting and
important \citep[e.g.,][]{hu,kneib04,pello}.  However, deliberate
selection of cluster fields makes the probability analysis completely
different from what we study here.} \citep[i.e., fields not
specifically chosen for the presence of a massive cluster lens;e.g.,][]
{rhoads00,rhoads01,steidel,bouwens,ouchi,stanway,pirzkal}.

This paper is organized as follows.  In \S~2, we review the relevant
lens theory and summarize our calculation methods.  In the technical
core of the paper (\S\S~3-4), we study the magnification properties
of simple but useful lens potentials: ellipsoidal isothermal and NFW
halos, with external tidal shear.  The idea is to identify general
features and understand the parameter dependences.  Building on this
foundation, in \S~5 we compute the probability of magnification without
multiple imaging for a realistic lens population, and discuss the
implications for the SDSS quasars.  Finally, in \S~6 we summarize our
results.  Throughout the paper, we assume a $\Lambda$CDM cosmology with
$\Omega_M = 0.3$, $\Omega_\Lambda = 0.7$, $\sigma_8 = 0.9$,
and $H_0 = 70$~km/s/Mpc, consistent with the recent results from WMAP
\citep{spergel03}.

\section{Computation Methods}

The lensing properties of a system can be derived from the lens
potential $\phi$, which is given by the solution to the 2-d Poisson
equation $\nabla^2\phi = 2\kappa$.  Here
$\kappa = \Sigma/\Sigma_{\rm crit}$ is the surface mass density in
units of the critical density for lensing,
$\Sigma_{\rm crit} = (c^2 D_{os})/(4\pi G D_{ol} D_{ls})$, where
$D_{ol}$, $D_{os}$, and $D_{ls}$ are angular diameter distances
between the observer, lens, and source.  (See \citealt*{SEF} for a
full discussion of lens theory.)  The relation between the position
$\xx$ of an image and the position $\uu$ of the corresponding source
is given by the lens equation,
\begin{equation}
  \uu = \xx - {\bf \nabla}\phi(\xx)\,.
\end{equation}
The magnification of an image at position $\xx$ is
\begin{equation}
  \mu(\xx) = \left| \begin{array}{cc}
    1 - \frac{\partial^2\phi}{\partial x^2} &
      - \frac{\partial^2\phi}{\partial x \partial y} \\
      - \frac{\partial^2\phi}{\partial x \partial y} &
    1 - \frac{\partial^2\phi}{\partial y^2}
  \end{array} \right|^{-1}
\end{equation}
The lensing critical curves are curves in the image plane where the
magnification is formally infinite, and the caustics are the
corresponding curves in the source plane.  Sources that lie outside
the caustics are singly imaged, while sources inside the caustics
have multiple images.  We use the {\em gravlens} software by
\citet{lenscode} to find the caustics, solve the lens equation, and
compute the image magnifications for the various models we consider.

We are interested in the cross section for having a magnification
(of a single image, or a combination of unresolved images) larger
than $\mu$, which we compute with Monte Carlo simulations.  For
singly-imaged sources, we set a minimum magnification $\mumin$ of
interest, typically $\mumin = 1.5$, and find the smallest circle
in the image plane such that all images outside the circle have
$\mu < \mumin$; this ensures that all of the images of interest lie
inside the circle.  We then pick $\sim\!10^{6}$ random image positions
in the circle.  The cross section for producing a singly-imaged system
with magnification greater than $\mu$ can be written as
\begin{equation}
  A_{\rm sing}(\mu) = \int_{\mu(\uu)>\mu} d\uu
  = \int_{\mu(\xx)>\mu} \frac{1}{\mu(\xx)}\ d\xx\,.
\end{equation}
The first integral is over all source positions $\uu$ where there
is only one image and it has magnification greater than $\mu$.  The
second integral is over the corresponding image positions, and the
equality holds because $\mu^{-1} = |\partial\uu/\partial\xx|$ is the
Jacobian of the transformation between the image and source planes.
In other words, the cumulative singly-imaged magnification
distribution can be computed by simply summing the images, weighted
by their inverse magnifications.

For multiply-imaged sources, we find the smallest circle enclosing
the caustics, then pick $\sim\!10^{6}$ random sources in this circle
and solve the lens equation to find the image configurations.  Now
we throw away the singly-imaged systems, and use the multiply-imaged
systems to compute the cross section for having a magnification greater
than $\mu$ but no extra {\em detectable} images,
\begin{equation}
  A_{\rm mult}(\mu) = \int_{\mu(\uu)>\mu} S(\uu)\,d\uu\,.
\end{equation}
This integral spans the multiply-imaged region, but the function
$S(\uu)$ selects source positions that produce lenses with only a
single detectable image; specifically, $S(\uu)$ is 1 if the additional
images are undetectable (either because they are too faint, or too
close to the brightest image), and 0 otherwise.  In general, $S(\uu)$
will depend on the specific source, instrument, and observational
conditions.  Hereafter we refer to it as the ``single-detectable-image
criterion'' (SDIC).  In recent HST images of the $z \approx 6$ SDSS
quasars, \citet{richards} were able to rule out the presence of extra
images with a flux ratio relative to the quasar of $f>0.01$ down to a
separation $\Dt>0\farcs3$, or brighter than $f>0.1$ down to
$\Dt>0\farcs1$.  We consider both of these SDICs in our analysis.
In configurations where there are multiple images that would not be
resolved, we include all of them in the net magnification.

\section{Isothermal Halos}

The isothermal ellipsoid is a simple but surprisingly useful model
for studying lensing by galaxies.  In this section we delineate the
situations in which isothermal halos can produce magnification
without detectable multiple images.

\subsection{Definitions of Ellipticity and Shear}

Early-type galaxies, which dominate the lensing optical depth at image
separations $\Dt \lesssim 4\arcsec$, appear to have nearly-isothermal
profiles based on evidence from strong and weak lensing, stellar dynamics,
satellite kinematics, and X-ray studies \citep[e.g.,][]{fabbiano,zaritsky,
rix97,gerhard,mckay,tk2016,kt1608,rkk,sheldon}.  The 3-d density
$\rho \propto r^{-2}$ and projected surface mass density
$\kappa \propto R^{-1}$ correspond to a flat rotation curve or velocity
dispersion profile, and a deflection angle that is independent of impact
parameter.

The lensing properties of a {\em singular isothermal sphere} are very
simple \citep[e.g.,][]{SEF}.  The surface mass density is
$\kappa = \Rein/(2R)$ where $\Rein$ is the Einstein radius, and the
lensing potential is $\phi = \Rein\,R$.  A source at radius $u>\Rein$
behind the lens produces a single image at radius $R = \Rein + u$ which
has magnification $\mu = R/(R-\Rein) = 1+\Rein/u$.  A source at radius
$u<\Rein$ produces two images at radii $R_{\pm} = \Rein \pm u$ on
opposite sides of the lens galaxy, which have magnifications
$\mu_{\pm} = R_{\pm}/(R_{\pm}-\Rein) = 1 \pm \Rein/u$ (where a
negative magnification means that the image is parity reversed).

An {\em isothermal ellipsoid} has a projected surface mass density of
\begin{equation} \label{eq:sie}
  \kappa(R,\theta) = \frac{b}{2R} \left[\frac{1+q^2}
    {(1+q^2)-(1-q^2)\cos2\theta}\right]^{1/2} ,
\end{equation}
where $q \le 1$ is the axis ratio, so the ellipticity is $e=1-q$,
and $(R,\theta)$ are polar coordinates centered on the lens galaxy.
(Without loss of generality, we are working in coordinates aligned
with the major axis of the galaxy.)  The lensing properties of an
isothermal ellipsoid are given by \citet{kassiola}, \citet*{kormann},
and \citet{spirals}.  For a spherical galaxy the parameter $b$
equals the Einstein radius, while for a nonspherical galaxy we
can relate them by \citep[see][]{huterer}
\begin{equation}
  \frac{\Rein}{b} = \frac{1}{\pi}\
    \left[2(1+q^{-2})\right]^{1/2}\ K\left(1+q^{-2}\right)\,,
\end{equation}
where $K(x)$ is the elliptic integral of the first kind.

{\em Gravitational tidal shear}, produced by objects near the main
lens halo or projected along the line of sight, can increase the
probability for high magnifications.  Shear is expected to be common,
based on both analytic estimates and numerical simulations
\citep[e.g.,][]{KKS,holder}, and it is generally required for fitting
observed galaxy-mass ($M \sim 10^{12}\,M_\odot$) strong lens systems
\citep[e.g.,][]{KKS,witt}.  The lens potential associated with shear
is
\begin{equation}
  \phi(R,\theta) = - \frac{1}{2}\,R^2\,\gamma\,\cos2(\theta-\theta_\gamma)\,,
\end{equation}
where $\gamma$ is the dimensionless shear amplitude,
$\theta_\gamma$ is the shear direction.  Shears of
$\gamma \sim 0.05$--$0.1$ are common for galaxy-mass lenses, and
shears of $\gamma \sim 0.2$--$0.3$ are possible for lens galaxies
lying in dense environments \citep[e.g.,][]{KKS,witt,kundic1115,
kundic1422,fischer,kneib,holder}.

\subsection{Parameter Dependences for a Single Lens}

\begin{figure}
\centerline{\epsfxsize=3.1in \epsfbox{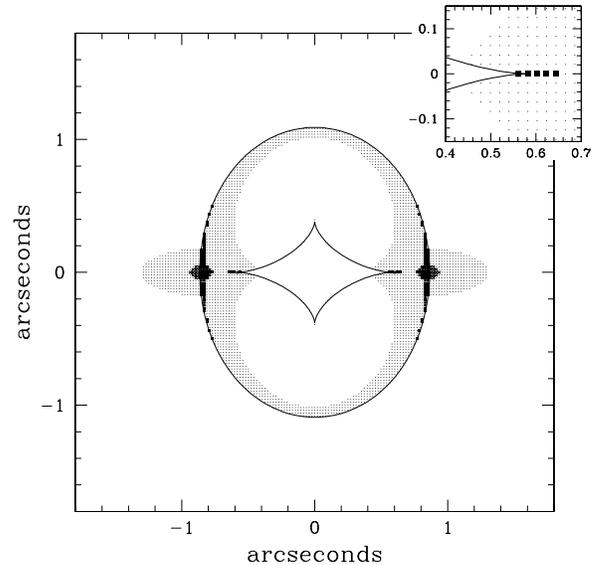}}
\caption{
Source plane for an isothermal ellipsoid with ellipticity $e=0.5$
and Einstein radius $\Rein=1\farcs0$.  The curves show the caustics.
The small (large) points outside the caustics indicate singly-imaged
sources with magnification $\mu>3$ ($\mu>5$).  The points inside the
caustic show multiply-imaged sources for which extra images are
undetectable; the small (large) points denote a flux ratio
threshold $f<0.1$ ($f<0.01$).  The inset shows a close-up of the
tip of the inner caustic.
}\label{fig:toy-sie}
\end{figure}

To begin to understand isothermal lenses, we show the source plane for
a sample lens with ellipticity $e=0.5$ in \reffig{toy-sie}.
Singly-imaged sources with magnification $\mu>3$ occur only in a
region just outside the caustics and near the minor axis of the radial
caustic (which corresponds to the major axis of the galaxy density
distribution).  In this example with Einstein radius $\Rein=1\arcsec$,
the image separations are larger than HST resolution, so the only
systems that have only one detectable image are those whose extra
images are too faint.  Most of these are small flux ratio doubles,
corresponding to sources that lie in two regions: just inside the
outer radial caustic, where the secondary image is very faint; or near
the inner tangential caustic, especially near the cusps along the
major axis, where the primary image is highly magnified.  In this
example the two regions merge together when the flux ratio
threshold for missing the second image is $f<0.1$ (the small
points in \reffig{toy-sie}), but remain distinct when the
threshold is $f<0.01$ (the large points).

\begin{figure}
\centerline{\epsfxsize=3.1in \epsfbox{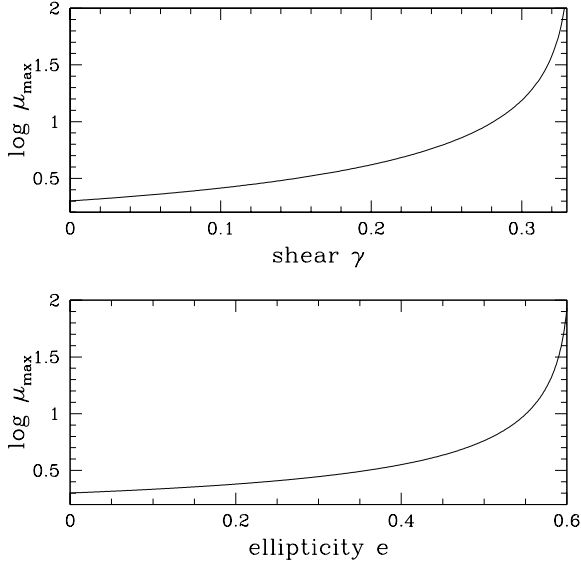}}
\caption{
Maximum singly-imaged magnification for an isothermal sphere with
shear (upper panel) and for an isothermal ellipsoid (lower panel).
The maximum magnification becomes infinite at $\gamma=1/3$ or
$e=0.606$ (see the Appendix for details).
}\label{fig:iso-max}
\end{figure}

\begin{figure*}
\centerline{\epsfxsize=6.2in \epsfbox{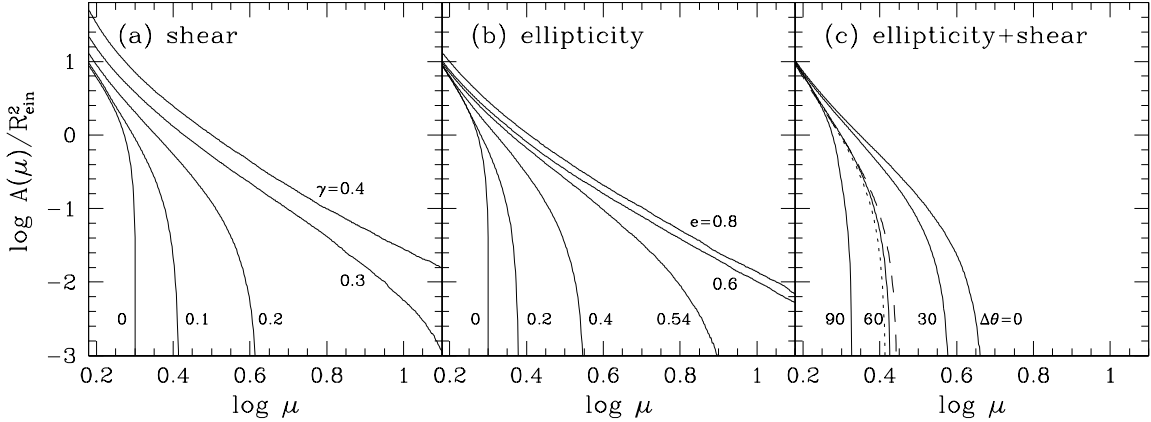}}
\caption{
Singly-imaged magnification distributions for isothermal models.
The area $A(\mu)$ where the magnification is greater than $\mu$ is
expressed in units of $\Rein^2$.
(a) Effects of shear.
(b) Effects of ellipticity.
(c) Effects of ellipticity and shear together; the ellipticity and
shear are fixed at $e=0.3$ and $\gamma=0.1$, and we vary the angle
between them.  For reference, the dashed curve shows a model with
$e=0.3$ and no shear, while the dotted curve shows a model with
$\gamma=0.1$ and no ellipticity.
}\label{fig:iso}
\end{figure*}

\begin{figure*}
\centerline{\epsfxsize=6.2in \epsfbox{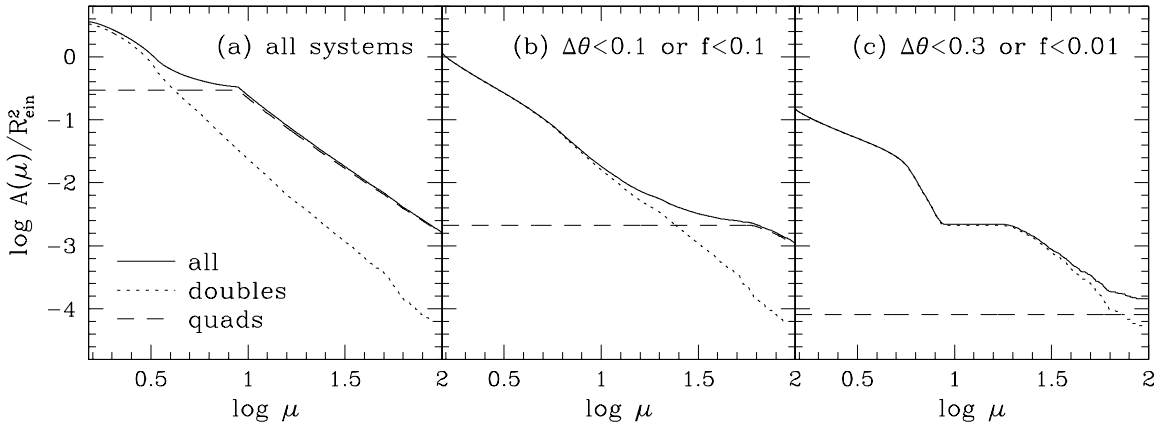}}
\caption{
Multiply-imaged magnification distributions for an isothermal
ellipsoid with ellipticity $e=0.5$.  The different line types denote
different image configurations (doubles, quads, or all lenses).  The
different panels show different criteria for having only a single
detectable image; panels (b) and (c) correspond to the detection
limits in HST observations of the $z \approx 6$ SDSS quasars by
\citet{richards}.
}\label{fig:iso-dist1}
\end{figure*}

\begin{figure*}
\centerline{\epsfxsize=6.2in \epsfbox{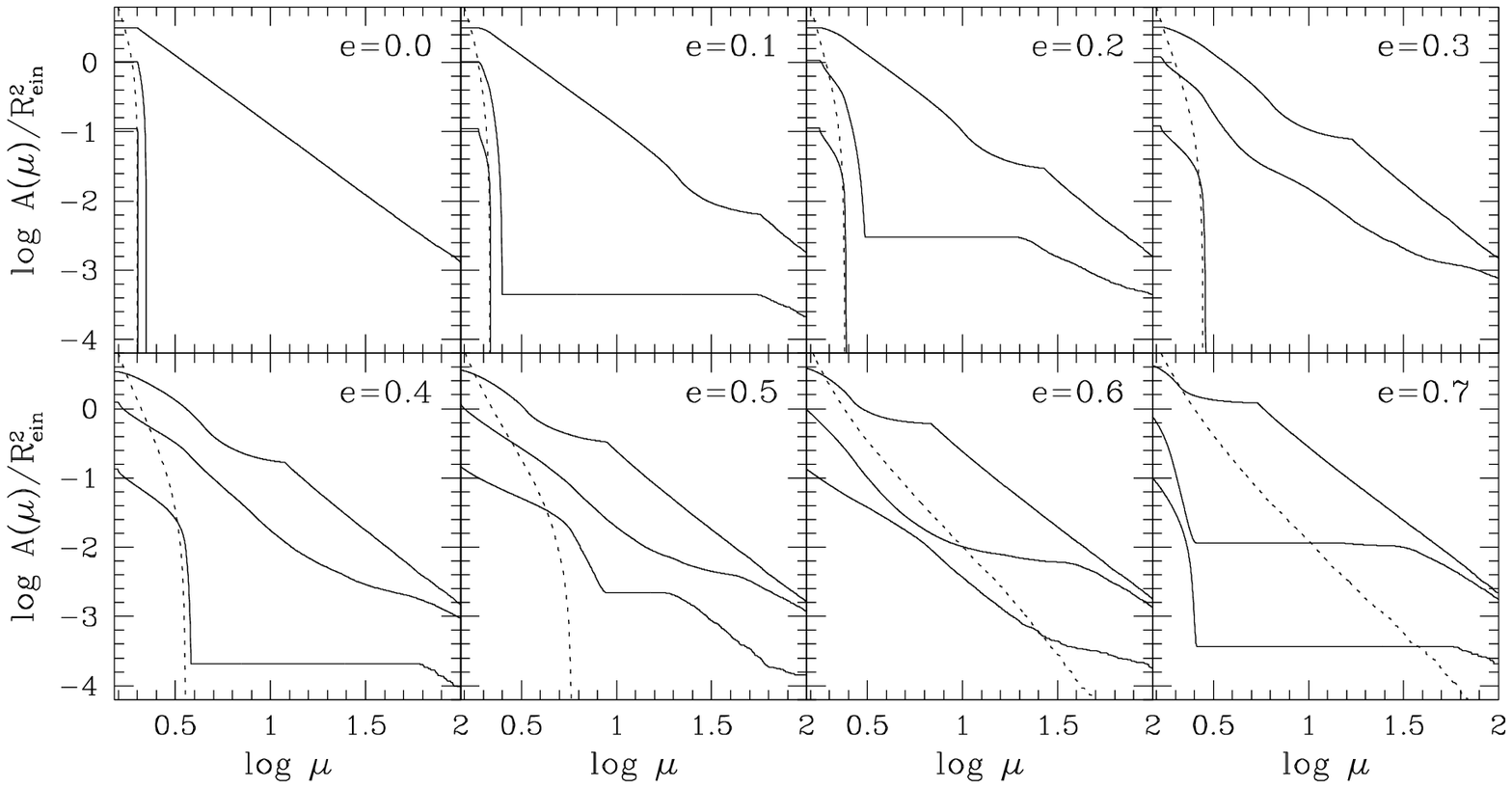}}
\caption{
Effects of ellipticity on multiply-imaged magnification distributions
for isothermal models.  In each panel, the three solid curves show
different criteria for having a single detectable image; the
top curve shows all systems, the middle curve shows $\Dt<0\farcs1$
or $f<0.1$, and the bottom curve shows $\Dt<0\farcs3$ or $f<0.01$.
For comparison, the dotted curve shows the appropriate singly-imaged
magnification distribution (see \reffig{iso}).
}\label{fig:iso-dist2e}
\end{figure*}

We see that for isothermal galaxies with $e=0.5$, the cross section
for magnified systems with undetectable extra images appears to be
larger than the cross section for magnified singly-imaged systems.
In the next two subsections we quantify this result carefully by
computing the magnification cross sections for both cases.

\subsubsection{Singly-imaged magnification distributions}

For most isothermal lenses (unless the ellipticity or shear is
large), there is a finite upper bound on the singly-imaged
magnification.  The bound is derived in the Appendix, and shown
in \reffig{iso-max}.  While spherical models can never produce a
singly-imaged magnification larger than $\mumax=2$ (in the absence
of microlensing; see \citealt{WL02}), non-spherical models can in
principle produce much larger magnifications.  However, the effect
is not likely to be very dramatic in practice: for a typical shear
$\gamma \sim 0.1$ or ellipticity $e \sim 0.3$, $\mumax$ is still
less than 3.

The full magnification cross sections are shown in \reffig{iso}
for various values of the parameters.  Since all of the physical
parameters --- the lens galaxy mass and redshift, and the source
redshift --- are contained in the Einstein radius, the dimensionless
cross section $A(\mu)/\Rein^2$ depends only on ellipticity and shear
so the parameter space we must study is small.  Panels (a) and (b)
show that shear and ellipticity increase not just $\mumax$ but the
whole high-magnification tail.  However, the cross section for high
magnifications is small; even in models with large shear or ellipticity,
the area in the source plane with $\mu>10$ is less than 0.1\% of the
area with $\mu>1.5$.

Shear and ellipticity are not mutually exclusive, so panel (c)
shows what happens when we include both; we fix the amplitudes to
typical values $\gamma=0.1$ and $e=0.3$ and vary the angle between
them.  The effects are largest when the shear and ellipticity are
aligned, because they combine to increase the quadrupole moment of
the lens potential; and the effects are smallest when they are
orthogonal because their quadrupoles partially cancel.  When
averaged over angle, we expect the combination of shear and
ellipticity to produce a modest increase in the area with modest
magnifications.

\subsubsection{Multiply-imaged magnification distributions}

Sample multiply-imaged magnification distributions are shown in
\reffig{iso-dist1}.  There are various unusual features that can be
understood with the help of the corresponding source plane shown in
\reffig{toy-sie}.  Without imposing any single-detectable-image
criterion (SDIC), the magnification cross section shows a kink at the
minimum magnification for quads, and the high-magnification systems
are dominated by quads \citep[see, e.g.,][]{BK87,SEF,finch}.
SDICs remove the vast majority of quads, however, because the
extra images in quads tend to be fairly bright.  The only quad sources
that survive the cut lie extremely close to the caustic, with image
configurations dominated by a very bright and very close pair of
images.  In this example, when the flux ratio SDIC is $f<0.1$
the magnification distribution is a smooth curve.  However, when the
flux ratio SDIC is $f<0.01$ the distribution breaks up into two
separate populations, with the low-magnification population lying just
inside the radial caustic, while the high-magnification population
lies just outside the cusps of the tangential caustic (see
\reffig{toy-sie}).  (The image separation SDIC is unimportant
here because the Einstein radius is larger than the HST resolution.)

Having understood the general features, we can now examine how the
multiply-imaged magnification distribution depends on ellipticity,
as shown in \reffig{iso-dist2e}.  In the absence of SDICs,
ellipticity raises the high-magnification tail; for example, the
cross section for $\mu>10$ is increased by a factor of $\sim$2 for
$e \gtrsim 0.5$.  This case will apply to low-mass halos where the
Einstein radius is small enough that all image separations are
unresolvable.  When the flux ratio SDIC is important (when the
halo mass is large enough that the image separations would be
resolved), ellipticity has a much more dramatic effect.  In the
spherical case there are no magnifications $\mu > 2/(1-f)$, where $f$
is from the SDIC.  Introducing ellipticity creates a population
of high-magnification sources lying just outside the cusps of the
tangential caustic.  As $e$ increases, that population grows and
merges with the population of lower-magnification sources lying inside
the radial caustic (as in \reffig{iso-dist1}b).  Finally, as the
ellipticity grows to $e>0.606$ the cusp of the tangential caustic
pierces the radial caustic, so the high-magnification region just
outside the cusp becomes associated with singly-imaged rather than
multiply-imaged systems.  This explains why the multiply-imaged cross
section curve changes shape, but it is not very important in practice
because such large ellipticities are rare.

The details clearly depend on the ellipticity and the SDIC, but
the most important result is more general: the ordering of the curves
in \reffig{iso-dist2e}.  The dominant source of magnifications $\mu
\gtrsim 5$ should be unresolvable small-separation lenses produced by
low-mass halos, followed by systems with multiple images where the
extra images are too faint to be detected.  The contribution from true
singly-imaged systems is generally not as important, except when the
ellipticity is large ($e \gtrsim 0.6$), and even then the cross
section is quite small.

Shear has similar effects on the magnification cross sections,
because like ellipticity it increases the quadrupole moment of the
lens potential and makes the tangential caustic larger.  We do not
show the cross sections for different shears, because the results 
appear very similar to those displayed in \reffig{iso-dist2e}.

\subsection{Averaging over Ellipticity and Shear}
\label{sec:iso-avg}

In order to obtain the overall probability distribution for the
magnification, we next average over realistic distributions of
ellipticity and shear.  For the distribution of shear amplitudes,
we use the model derived by \citet{holder} for the environments
of early-type galaxies in $N$-body and semi-analytic models of
galaxy formation; they find a lognormal distribution with median
$\gamma=0.05$ and dispersion $\sigma_\gamma = 0.2$~dex.  We use
random shear directions.  For the ellipticity distribution, we use
data on the shapes of observed early-type galaxies.\footnote{The
data give the shape of the light distribution, while what we need
is the shape of the mass distribution.  The mass and light shapes
may not be correlated on a case-by-case basis, but for our purposes
it is sufficient to assume that their distributions are similar
\citep[see][]{rusin}.}  \citet*{jorgensen} give ellipticities for
379 E and S0 galaxies in 11 clusters, including Coma.  The
distribution is broad, with mean ${\bar e} = 0.31$ and dispersion
$\sigma_e = 0.18$.  We average over more than 1000 random
combinations of ellipticity and shear.

\begin{figure}
\centerline{\epsfxsize=3.1in \epsfbox{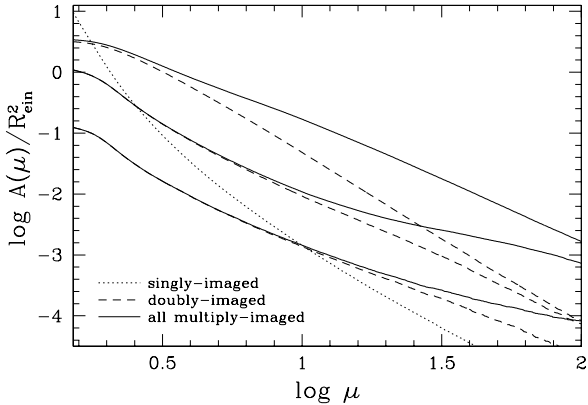}}
\caption{
Magnification cross sections after averaging over ellipticity and shear.
The dotted curve shows the singly-imaged cross section.  The solid and
dashed curves show the multiply-imaged cross sections.  The top pair of
curves includes all multiply-imaged systems (regardless of the
number of detectable images).  The middle pair corresponds to the
following criteria for having a single detectable image: $\Dt<0.1\,\Rein$
or $f<0.1$; while the bottom pair corresponds the criteria $\Dt<0.3\,\Rein$
or $f<0.01$.
}\label{fig:iso-edist}
\end{figure}

\reffig{iso-edist} shows the resulting cross sections.  Singly-imaged
systems are important only for low magnifications ($\mu\lesssim 2$).
If the Einstein radius is small and the lenses are unresolved, then
multiply-imaged systems dominate at $\mu \gtrsim 1.8$; much of the
relevant cross section comes from quads (as indicated by the
difference between the solid and dashed curves).  When the flux ratio
SDIC applies, multiply-imaged systems dominate at $\mu
\gtrsim 2.5$ for a SDIC of $f<0.1$; for $\mu \lesssim
10$ most of this cross section comes from doubles, while for higher
magnifications there is a significant contribution from quads.  For a
flux ratio SDIC $f<0.01$, multiply-imaged systems
dominate at $\mu \gtrsim 10$, and most of the cross section comes from
doubles.  The overall conclusion is that for isothermal halos, most
large magnifications $\mu \gtrsim 10$ will correspond to
multiply-imaged systems where the extra image are not detectable
(either unresolved or faint).

We have explicitly examined ellipticity and shear, but early-type
galaxies are also observed to have small octopole moments in their
light distributions.  We have repeated our analysis using the
ellipticity and octopole distributions from the galaxy samples of
\citet{bender} and \citet*{saglia}, and confirmed that our results
are not very sensitive to changes in the ellipticity distribution
or to the addition of octopole terms.

\section{NFW Halos}

Another common and useful lens model is the Navarro-Frenk-White
(NFW) profile, which describes halos produced in $N$-body simulations.
The NFW profile is thought to describe systems that are dominated by
dark matter at all radii: massive cluster halos, and perhaps low-mass
dwarf halos as well.  In this section we study the ability of NFW
lenses to produce magnification without detectable multiple images.

\subsection{Definitions}

The NFW profile has the form
\begin{equation}
  \rho(r) = \frac{\rho_s}{(r/r_s)(1+r/r_s)^2}\ ,
\end{equation}
where $r_s$ is a scale radius and $\rho_s$ is a characteristic
density.  There has been debate about whether the inner density
profile of simulated clusters really asymptotes to the
$\rho \propto r^{-1}$ form \citep[e.g.,][]{navarro97,fukushige97,
moore99,jing00,power03,fukushige04}, and whether such a density
cusp is consistent with observed clusters \citep[e.g.,][]{tyson,
smith,ettori,kelson,sand02,lewis,sand04}.  Our main need is for
a model other than the isothermal ellipsoid that we can apply
to massive clusters and low-mass dwarf halos.  For this purpose
the NFW model is standard and sufficient, and exploring a larger
family of models (such as generalized NFW) is beyond the scope
of this paper.  Besides, with appropriate normalizations
generalized NFW profiles lead to lens statistics that are not
so sensitive to the inner profile slope \citep{KM01}.

The NFW profile has projected surface mass density \citep{bartelmann}
\begin{equation}
  \kappa(R) = 2\,\kappa_s\ \frac{1-F(R/r_s)}{(R/r_s)^2-1}\ ,
\end{equation}
where $\kappa_s \equiv \rho_s\,r_s/\Sigma_{\rm crit}$ is a
dimensionless lensing ``strength'' parameter, and the function
$F(x)$ is:
\begin{equation}
  F(x) = \cases{
    (1-x^2)^{-1/2}\ \mbox{tanh}^{-1} (1-x^2)^{1/2} & $x<1$ \cr
    1 & $x=1$ \cr
    (x^2-1)^{-1/2}\ \mbox{tan}^{-1} (x^2-1)^{1/2} & $x>1$
  }
\end{equation}
We obtain an elliptical NFW model by replacing
$R \to (x^2+y^2/q^2)^{1/2}$ in the surface mass density, where
$q \le 1$ is the projected axis ratio, and the ellipticity is
$e = 1-q$.  The lensing properties of an elliptical NFW model
can be computed with a set of 1-d numerical integrals
\citep{schramm,catalog}.\footnote{It is possible to obtain an
analytic NFW model by putting the elliptical symmetry in the
potential rather than the density \citep*[e.g.,][]{golse,meneghetti}.
However, $N$-body simulations suggest that it is the density rather
than the potential that has ellipsoidal symmetry (or more generally,
triaxiality; e.g., \citealt{JS02}).  We find that working with an
elliptical density and using numerical integrals is not a major
hindrance.}

\begin{figure}
\centerline{\epsfxsize=3.1in \epsfbox{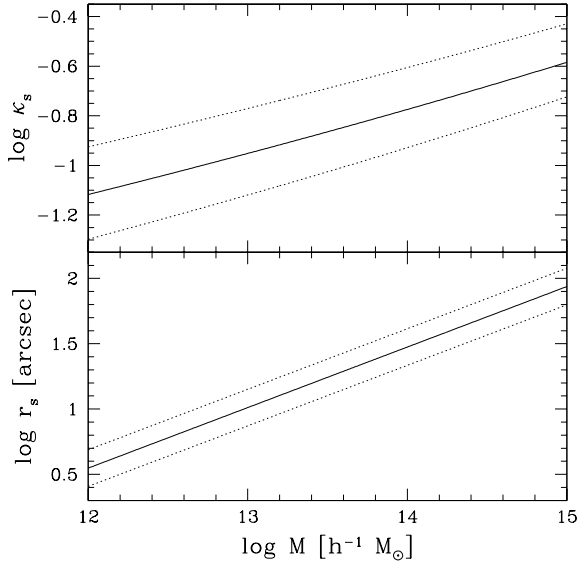}}
\caption{
Sample lensing strength $\kappa_s$ (upper panel) and halo scale radius $r_s$
(lower panel) versus halo mass, for a lens at redshift $z_l=1$ and source
at redshift $z_s=6$.  The solid curves show results for halos with
the median concentration, using the median $c(M)$ relation from
\citet{bullock}; the dotted curves indicate the 1$\sigma$ range due
to the scatter in the $M$-$c$ correlation, $\sigma_c=0.14$~dex at
fixed mass.
}\label{fig:NFWnorm}
\end{figure}

\begin{figure*}
\centerline{
  \epsfxsize=3.1in \epsfbox{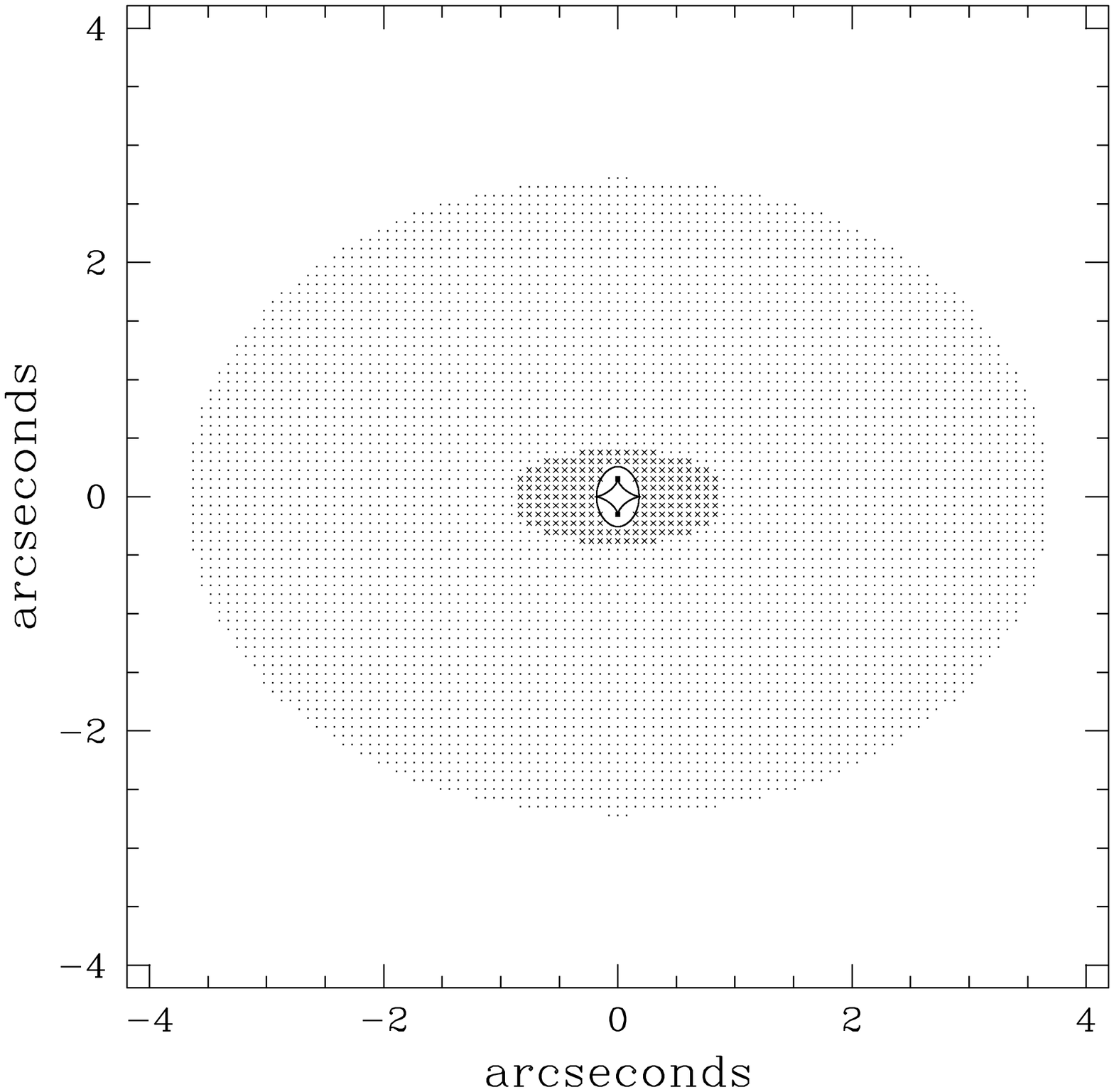}
  \epsfxsize=3.1in \epsfbox{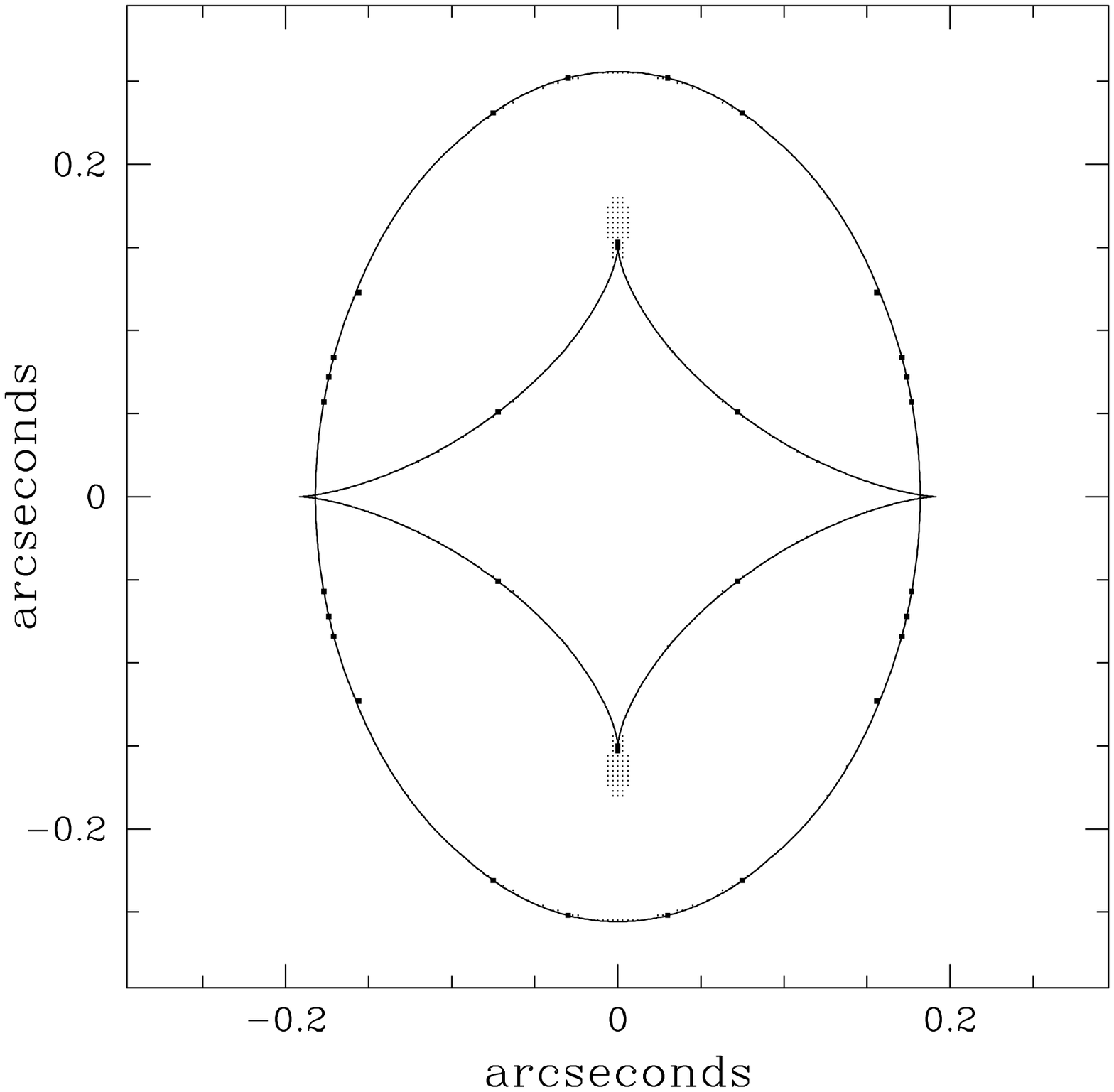}
}
\caption{
Source plane for an NFW halo with strength $\kappa_s=0.168$, scale
radius $r_s = 167\,h^{-1}\mbox{ kpc} = 29\farcs8$, and ellipticity
$e=0.1$.  The curves show the caustics.  The small (large) points
outside the caustics indicate singly-imaged sources with magnification
$\mu>3$ ($\mu>10$).  The small points inside the caustic show
multiply-imaged sources for which extra images are undetectable;
the criteria for having only a single detectable image are
an image separation $\Dt<0\farcs1$ or flux ratio $f<0.1$ for the
small points, and $\Dt<0\farcs3$ or $f<0.01$ for the large points.
The left panel shows a large region of the source plane, while the
right panel shows a close-up of the multiply-imaged region.
}\label{fig:toy-nfw}
\end{figure*}

NFW profiles appear to form a two-parameter family specified by
$\rho_s$ and $r_s$, or equivalently by the virial mass $M$ and a
concentration parameter $c = r_{\rm vir}/r_s$.  (The virial radius
$r_{\rm vir}$ can be given as an explicit function of $M$ and $c$.)
In fact, the two parameters are correlated, and different models for
median relation and scatter have been proposed
\citep[e.g.,][]{navarro97,eke,bullock,JS02}.  For our purposes, the
important result is that the scale length $r_s$ and lensing strength
$\kappa_s$ are both correlated with the halo mass, as shown in
\reffig{NFWnorm}.  In this section we can express cross sections in
units of $r_s^2$ so we need not examine the $r_s$ dependence
explicitly; but we do need to examine the dependence on $\kappa_s$.

\subsection{Parameter Dependences for a Single Lens}

To begin to understand NFW lenses, in \reffig{toy-nfw} we show the
source plane for a sample lens with strength $\kappa_s=0.168$
(corresponding to a median $10^{14}\,M_\odot$ halo at $z_l=1$, see
\reffig{NFWnorm}) and ellipticity $e=0.1$.  The caustics are small,
and the region of multiply-imaged systems that satisfy reasonable
single-detectable-image criteria (SDIC) is smaller still.
The reason is that the image separations for massive halos are large,
so the important SDICs involve the flux ratios.  NFW halos
generally produce large magnifications, and the range of magnification
{\em ratios} is not very broad.  As a result, the only sources that
survive the SDICs lie very near the caustics, where one of the
images has a very large magnification.

By contrast, there is quite a large region of the source plane where
sources are singly-imaged but have large magnifications.  Thus, for
NFW halos it appears that singly-imaged systems will be more important
than multiply-imaged systems for producing large magnifications with
a single detectable image.  In the next two subsections we quantify
this result carefully.

\subsubsection{Singly-imaged magnification distributions}

\begin{figure}
\centerline{\epsfxsize=3.1in \epsfbox{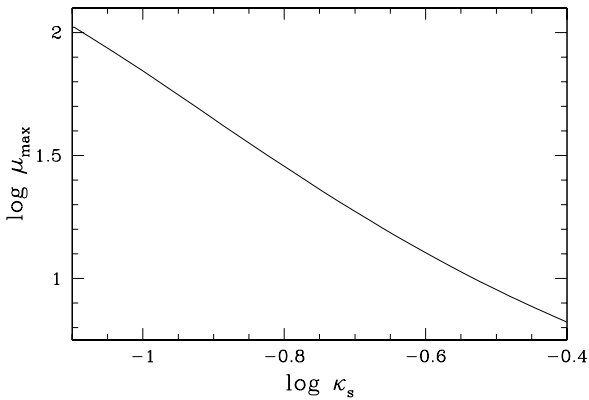}}
\caption{
Maximum singly-imaged magnification versus lensing strength for
spherical NFW lenses.
}\label{fig:nfw-max}
\end{figure}

Even spherical NFW lenses are complex systems where the lens
equation is transcendental, so they must be studied numerically.
\reffig{nfw-max} shows the maximum singly-imaged magnification
as a function of the strength $\kappa_s$.  Spherical NFW halos
can apparently produce large magnifications without multiple
imaging, especially when the lensing strength (or halo mass) is
small.  It has been known that for multiple imaging at fixed
splitting angle NFW lenses tend to produce smaller cross sections
but larger magnifications than isothermal lenses
\citep[see][]{BK87,knudson}.  Now we see that the association
between NFW lenses and high magnifications extends to single
imaging as well.

\begin{figure*}
\centerline{\epsfxsize=6.2in \epsfbox{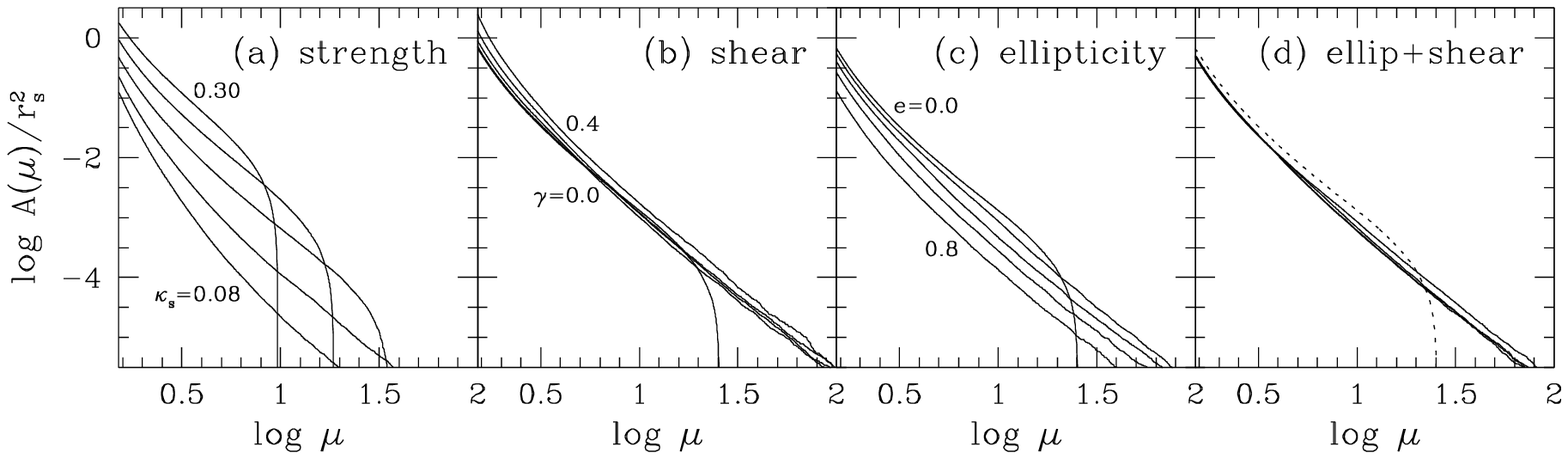}}
\caption{
Singly-imaged magnification distributions for NFW models.  The area
$A(\mu)$ where the magnification is greater than $\mu$ is expressed
in units of $r_s^2$ (not $\Rein^2$).
(a) Effects of the lensing strength $\kappa_s$; the curves show
$\kappa_s=0.08, 0.10, 0.14, 0.20, 0.30$ from bottom to top.
(b) Effects of shear; the curves show $\gamma=0, 0.1, 0.2, 0.3, 0.4$
from bottom to top.
(c) Effects of ellipticity; the curves show $e=0, 0.2, 0.4, 0.6, 0.8$
from top to bottom.
(d) Effects of ellipticity and shear together; the curves show
$\Dt = 0, 30, 60, 90$ deg.  The ellipticity and shear are fixed at
$e=0.3$ and $\gamma=0.1$.  For reference, the dotted curve shows a
spherical model.
In panels (b)-(d) the lensing strength is fixed at $\kappa_s=0.168$
(the median value for a $10^{14}\,h^{-1}\,M_\odot$ halo in
\reffig{NFWnorm}).
}\label{fig:nfw}
\end{figure*}

The full magnification distributions are shown in \reffig{nfw}.
Panel (a) shows that for spherical halos, increasing the strength
$\kappa_s$ decreases $\mumax$ but increases the overall cross
section.  Note that the figure shows the cross section in units
of $r_s^2$; expressing the area in physical units like steradians
would increase the distance between the curves, since the strength
is correlated with $r_s$ (see \reffig{NFWnorm}).

\begin{figure*}
\centerline{\epsfxsize=6.2in \epsfbox{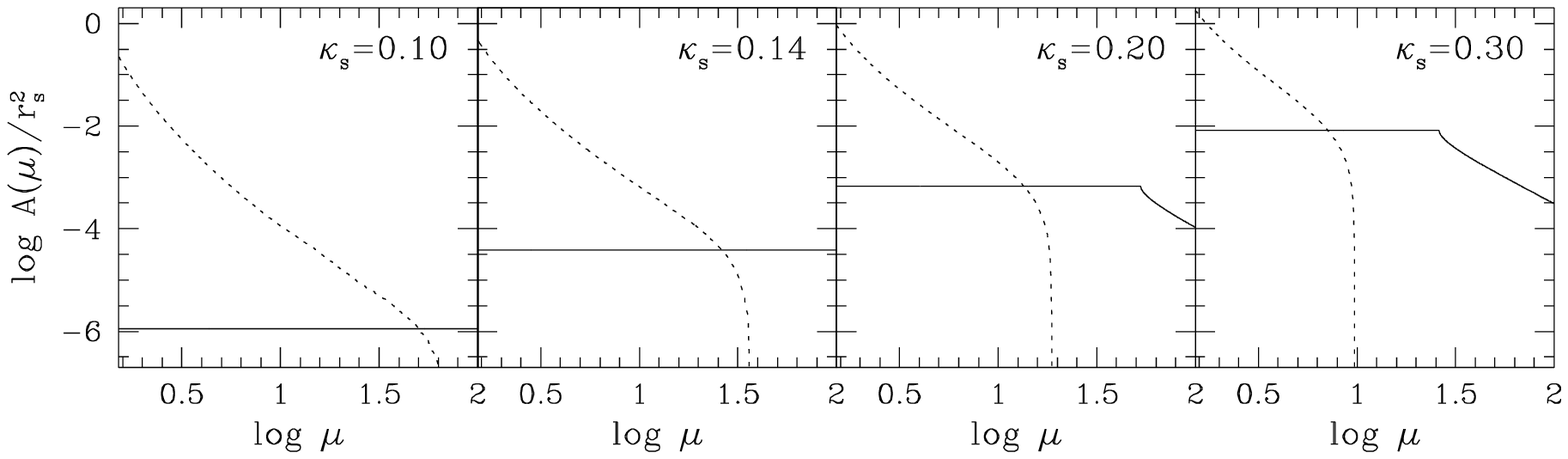}}
\caption{ Magnification distributions for spherical NFW halos with
different values of the lensing strength $\kappa_s$.  The solid curves
show multiply-imaged magnification distributions, regardless of
the number of detectable images.  For comparison, the dotted curves
show the corresponding singly-imaged magnification distributions from
\reffig{nfw}.  }\label{fig:nfw-dist2k}
\end{figure*}

Panels (b) and (c) in \reffig{nfw} show that shear and ellipticity
do not dramatically affect the magnification distributions for
NFW halos.  This stands in contrast to the case for isothermal
halos.  The difference is that the magnification distribution for
spherical NFW halos already extends to high magnifications, so any
increase due to ellipticity or shear is relatively more moderate.
The departure from spherical symmetry does raise the tail to very
high magnifications, but that effect is not very sensitive to
the degree of asymmetry (shear or ellipticity).  Interestingly,
ellipticity appears to {\em lower} slightly the cross section for
moderate magnifications.  Finally, \reffig{nfw}d shows that
allowing a combination of ellipticity and shear has little effect
other than extending the high-$\mu$ tail of the distribution.
While this is shown explicitly only for fixed $e=0.3$ and $\gamma=0.1$,
we expect this generic feature to hold for other combinations.

\subsubsection{Multiply-imaged magnification distributions}

\begin{figure}
\centerline{\epsfxsize=3.1in \epsfbox{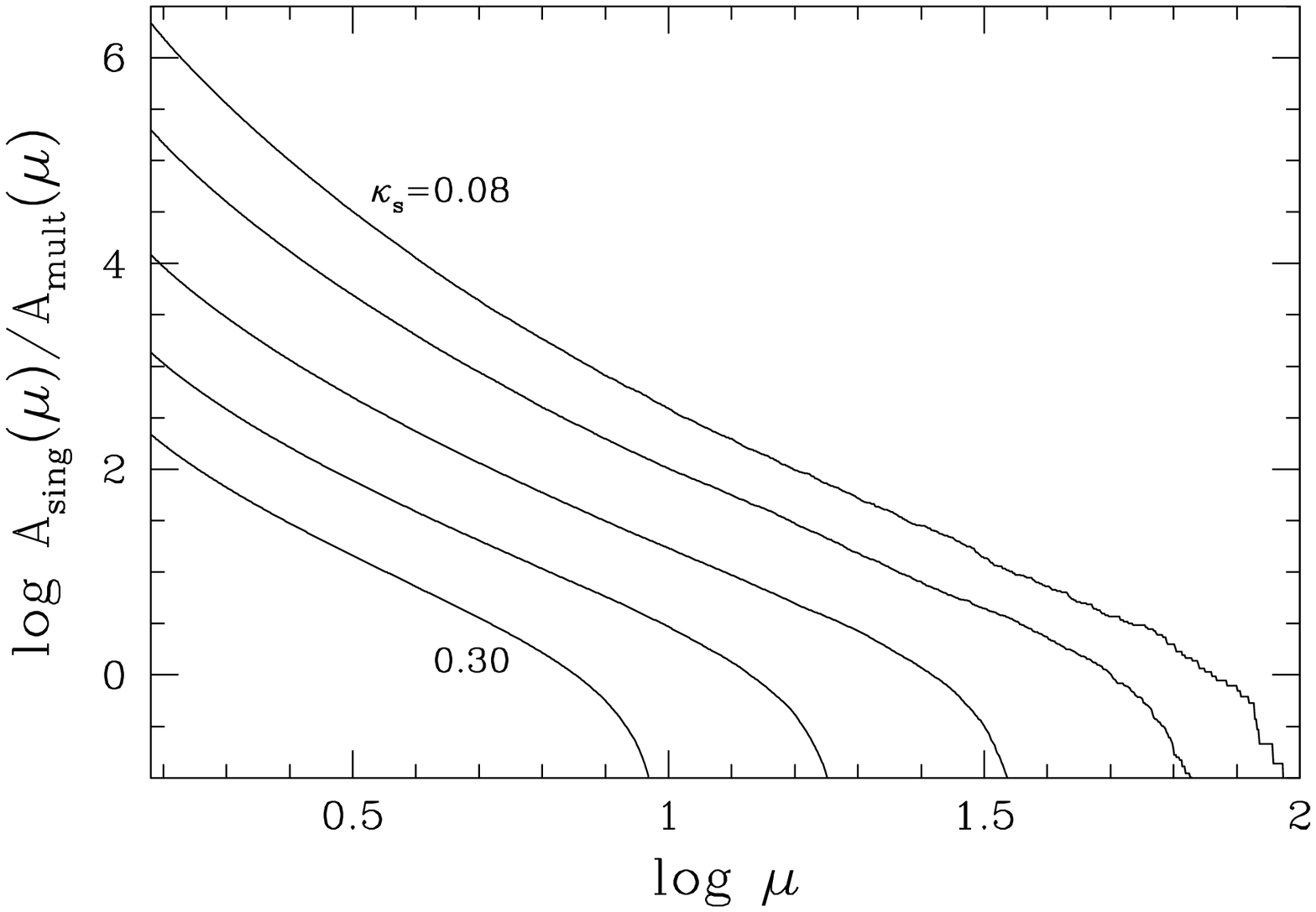}}
\caption{
Ratio of singly-imaged and multiply-imaged magnification cross
sections for NFW lenses.  The curves correspond to
$\kappa_s = 0.08, 0.10, 0.14, 0.20, 0.30$ from top to bottom.
}\label{fig:nfw2}
\end{figure}

\begin{figure*}
\centerline{\epsfxsize=6.2in \epsfbox{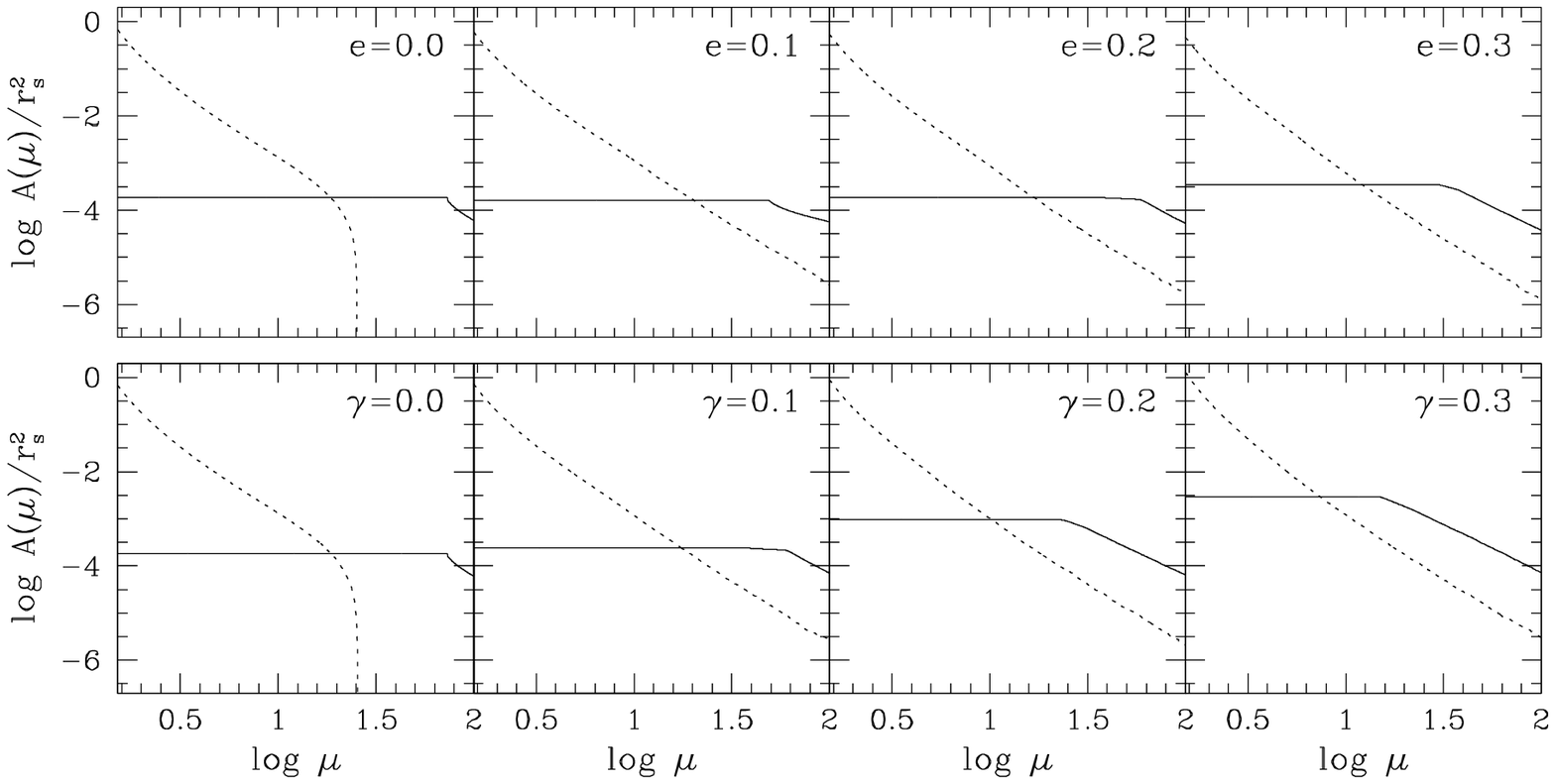}}
\caption{
Magnification distributions for NFW halos with different
ellipticities (upper panel) and shears (lower panel), with no
regard for the number of detectable images.  The lensing strength
is fixed at $\kappa_s=0.168$.  For comparison, the dotted curves
show the appropriate singly-imaged magnification distributions from
\reffig{nfw}.
}\label{fig:nfw-dist2eg}
\end{figure*}

\reffig{nfw-dist2k} shows sample multiply-imaged magnification
distributions for spherical NFW halos with different values of the
lensing strength $\kappa_s$.  Only the case with no regard for
the number of detectable images is shown, because as we saw in
\reffig{toy-nfw} the cross section for NFW lenses that satisfy the
flux SDICs is extremely small.  When the strength is low, the
cumulative cross section $A(\mu)$ is flat out to $\mu > 100$,
indicating that all magnifications are larger than 100; however, the
cross section is very small.  As $\kappa_s$ increases, the point at
which $A(\mu)$ begins to decline moves to the left, indicating that
the minimum magnification decreases; and the cross section increases.
For all but the most massive and concentrated halos, the
multiply-imaged magnification cross section is considerably smaller
than the singly-imaged cross section, except at the very highest (and
rarest) magnifications.

A better way to compare the singly- and multiply-imaged cases is
to take the ratio of the cross sections, as shown in \reffig{nfw2}.
Decreasing $\kappa_s$ increases the ratio dramatically, so for
moderate- to low-mass NFW halos the singly-imaged magnification
cross section can be orders of magnitude larger than the
multiply-imaged cross section.  This quantifies the statement
that for magnification by NFW halos, singly-imaged systems are
vastly more important than multiply-imaged systems.

\reffig{nfw-dist2eg} shows that ellipticity has little effect on
the multiply-imaged magnification distribution.  Large shears can
in principle increase the cross section.  However, it is not clear
that massive NFW halos could experience such large shears.  Shears
of $\gamma \sim 0.2$--$0.3$ typically occur when the lens is a
galaxy {\em within} a clusters, where the cluster serves as the
environment that produces the shear.  The filamentary structure
typical around clusters represents a very different environment,
for which the shear distribution is not well known.  Fortunately,
this uncertainty is not important for our results because shear
and ellipticity have such modest effects.  For simplicity, in the
rest of the paper we consider only spherical halos when computing
multiply-imaged magnification distributions for NFW halos.

\section{A Realistic Halo Population}

Having understood two fiducial halo models, we now combine
them into a realistic population of galaxies and clusters.  We
compute the overall probability for magnification with additional
detectable images, and use it to evaluate the hypothesis that the
four $z \approx 6$ SDSS quasars are highly amplified by lensing.
After defining the model (\refsec{popmod}), we present our general
results (\refsec{main}) and then apply them to the SDSS quasars
(\refsec{SDSS}).  We end with a discussion of some systematic
effects in our analysis (\refsec{systematics}).

\subsection{The Model}
\label{sec:popmod}

The total probability distribution for $\mu$ comes from integrating
the cross section over an appropriate halo population,
\begin{equation} \label{eq:P}
  P(\mu ; z_s) = \frac{1}{4\pi} \int dV \int dM\ \frac{dn}{dM}\
    A(\mu ; z_s, z_l, M)\,.
\end{equation}
The first integral is over the comoving volume between the observer
and source.  The second integral is over the comoving halo mass
function $dn/dM$; we adopt the theoretical mass function from
\citet{ST}.  Finally, $A(\mu)$ is the cross section computed above
(expressed in steradians), which depends on the source redshift $z_s$,
the lens redshift $z_l$, the lens halo mass $M$, and also on the lens
model (isothermal or NFW).  By using the appropriate cross section, we
can compute the probability for singly-imaged or multiply-imaged
magnifications by any of the halo populations we have considered.

Implicit in eqn.~(\ref{eq:P}) is an average over appropriate
ellipticity and shear distributions.  For isothermal halos we
use the results after averaging over ellipticity and shear, from
\refsec{iso-avg}.  Since the results for NFW halos are not very
sensitive to ellipticity and shear, we simply use a fixed
ellipticity $e=0.1$ for the singly-imaged case (in order to pick
up the high-magnification tail; see \reffig{nfw}), and we use
spherical models for the multiply-imaged case.

We consider a model with at least two different halo populations.
The most massive halos, corresponding to clusters and groups of
galaxies, are treated as NFW halos; while halos corresponding to
galaxies are modeled with isothermal profiles.  In this increasingly
standard model \citep{flores,keeton98,porciani,KW,li02}, the
difference between clusters and galaxies is usually attributed to
baryonic cooling: in massive halos the baryons have not had time
to cool so the systems retain their initial NFW form; while in
lower-mass halos the gas has cooled and condensed into the center
of the system, and created a more concentrated total (baryons +
dark matter) mass profile \citep[e.g.,][]{blumenthal,KW}.  Thus,
the transition between clusters and galaxies is characterized by
a mass scale $\Mclus$ such that halos with $M<\Mclus$ ($M>\Mclus$)
have a cooling time shorter (longer) than the age of the universe
and correspond to galaxies (clusters).

\begin{figure*}
\centerline{\epsfxsize=6.2in \epsfbox{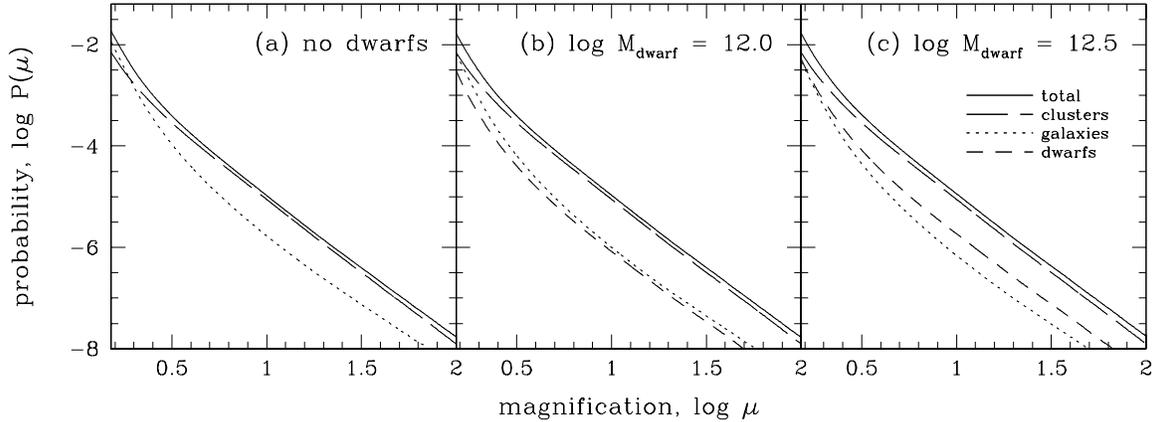}}
\caption{
Net singly-imaged magnification probability distributions for sources
at redshift $z_s=6$.  $P(\mu)$ is the cumulative probability of having
a magnification greater than $\mu$.  The solid curves show the total
probability, while the short-dashed, dotted, and long-dashed curves
show the contributions from dwarf halos, normal galaxies, and
clusters, respectively.  The different panels correspond to different
assumptions about the dwarf/galaxy transition (see text).
}\label{fig:pop1}
\end{figure*}

There may be a third population as well, namely low-mass ``dwarf''
halos with NFW profiles.  Theoretical halo mass functions rise much
more steeply than observed galaxy luminosity functions, leading to
the speculation that there may be a substantial population of
underluminous low-mass halos.  Mechanisms such as feedback or
reionization might have suppressed baryonic cooling and star
formation in low-mass systems \citep[e.g.,][]{dekel,efstat,ns97,
thoul,BKW,springel}, leading to halos that are dark and retain
their initial NFW form.  The apparent dearth of small-separation
lens systems implies a transition from isothermal galaxies back
to NFW dwarfs around a mass of $\Mdwarf \sim 10^{12}\,M_\odot$
\citep{li03,ma,kuhlen}.

Following \citet{ma}, we summarize the model by introducing a
function $f_{\rm SIS}(M)$ that describes the fraction of halos of
mass $M$ that have isothermal profiles (and the rest are NFW).
We use the function:
\begin{equation}
  f_{\rm SIS}(M) = \cases{
    0 & $M<\Mdwarf$ \cr
    1 & $\Mdwarf<M<\Mclus$ \cr
    \exp\left[-\frac{(\log M-\log \Mclus)^2}{\Sclus^2}\right] & $M>\Mclus$
  }
\end{equation}
To determine the parameter values, we follow \citet{kuhlen} and
fit the model to the observed image separation distribution from
the Cosmic Lens All-Sky Survey
\citep[CLASS;][]{myers,browne}.\footnote{Our quantitative results
differ slightly from the fiducial results of \citet{kuhlen}
because we now use the \citet{ST} mass function rather than that
from \citet{jenkins}, and we use $\sigma_8=0.90$ rather than
$\sigma_8=0.74$.}  The formal best-fit model has $\Sclus=0.13$
and $\log\Mclus=13.22$.  However, there is a degeneracy between
$\Sclus$ and $\Mclus$ such that a model with a sharp transition
($\Sclus=0$) and $\log\Mclus=13.38$ fits almost as well.  We have
verified that the two models give indistinguishable results, so
we report results only for the model with a sharp transition.  The
cluster/galaxy transition mass inferred from lensing agrees well
with estimates based on cooling arguments \citep[e.g.,][]{KW}.
As for the galaxy/dwarf transition, current data lack any ability
to determine whether it is smooth or sharp, so for simplicity we
use only a sharp transition.  The formal best-fit model has
$\log\Mdwarf=12.5$, although in fact only an upper limit on
$\Mdwarf$ is reliable \citep[see][]{kuhlen}.  This mass is too
high to be explained by reionization feedback, so it might indicate
either some other kind of feedback or some peculiarities in the
data (perhaps unknown incompletenesses at small separations, or
simply small-number statistics).  We therefore consider several
different values for $\Mdwarf$.  We take the transition masses
$\Mclus$ and $\Mdwarf$ to be independent of redshift.  This
assumption may seem objectionable, but we show below that the
specific value of $\Mclus$ has little effect on our results, and
we explicitly consider systematic uncertainties associated with
$\Mdwarf$.

\subsection{Main Results}
\label{sec:main}

\reffig{pop1} shows the net singly-imaged magnification
distribution, for our fiducial model.  For sources at redshift
$z_s = 6$, magnifications of $\mu \sim 1.6$ occur at the percent
level, and the distribution drops quickly.  The probabilities for
$\mu > (2,5,10)$ are $(0.3\%, 9\times10^{-5}, 1.1\times10^{-5})$.
In other words, significant singly-imaged magnifications are rare.

There are nevertheless several qualitative results that are
interesting and instructive.  First, for $\mu \gtrsim 1.8$ the
singly-imaged magnification probability is dominated by clusters.
This result seems at first glance to contradict \citet{CHS}, who
found that massive NFW halos had negligible impact on the probability
for lensing magnification, but the apparent discrepancy is easily
explained.  \citeauthor{CHS} considered lensing with undetected
multiple imaging, while we are considering the complementary case
where there is only a single image.  NFW halos are much more
efficient at producing highly magnified single images than multiple
images (see \reffig{nfw2}), which is why we find a much stronger
effect.  The important implication of this result is that if there
is a significant singly-imaged magnification ($\mu \gtrsim 5$)
then the lensing object is most likely a cluster, and ought to be
relatively easy to detect.  Conversely, lensing with small to
moderate singly-imaged magnifications ($\mu \lesssim 2$) is
dominated by galaxies.  This is consistent with the claim by
\citet{shioya} that one of the $z \approx 6$ quasars is magnified
by a factor $\mu \approx 2$ by a foreground galaxy, with no obvious
sign of a cluster.

\begin{figure*}
\centerline{\epsfxsize=6.2in \epsfbox{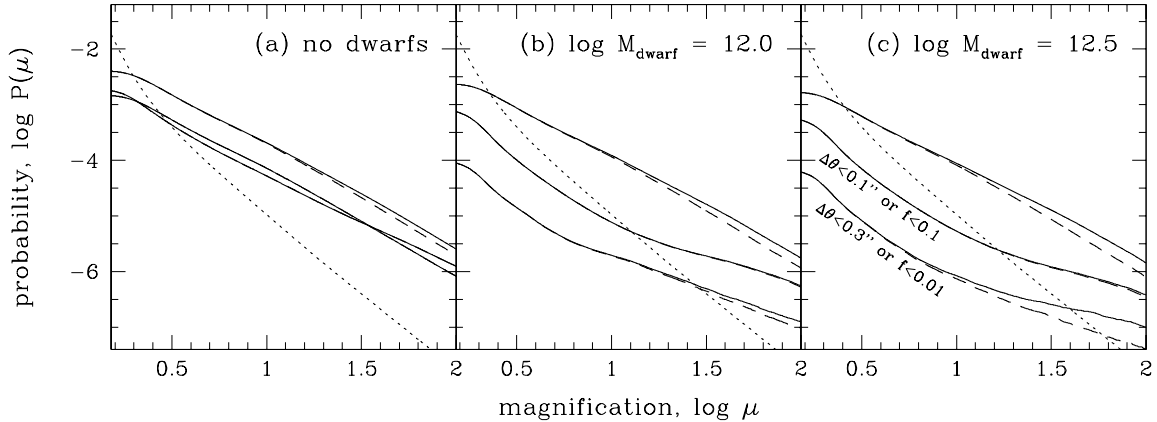}}
\caption{
Net multiply-imaged magnification probability distributions for
sources at redshift $z_s=6$.  In each panel, the upper solid curve
shows the overall multiply-imaged probability (regardless of the
number of detectable images), and the two other solid curves show the
probability with different criteria for detecting only a single
image (as indicated).  The dashed curves show the contribution to
each probability from galaxies alone (which is often indistinguishable
from the total).  The different panels correspond to different
assumptions about the dwarf/galaxy transition.  For comparison, the
dotted curve shows the net singly-imaged magnification distribution
from \reffig{pop1}.
}\label{fig:mult1}
\end{figure*}

A second interesting feature of the singly-imaged magnification
probability relates to the possible galaxy/dwarf transition at
low masses.  As discussed above, current lens data hint at the
transition but do not constrain it well, so we consider three
possible cases: (a) a model with no transition; (b) a model with
the transition at $\log\Mdwarf = 12.0$; and (c) a model with the
transition at $\log\Mdwarf = 12.5$, which perhaps seems high but
is formally the best fit to current data (although it is not
significantly better than the other models).  Models (a) and (c)
are extremes that bound the range of reasonable possibilities,
and model (b) is a sample intermediate model.  Not surprisingly,
\reffig{pop1} shows that moving the transition changes the relative
contributions of galaxies and dwarf halos to the probability; if
the transition occurs at $\log\Mdwarf \gtrsim 12.0$, dwarfs can
conceivably contribute more of the probability than normal
galaxies.  But the important result is that changing $\Mdwarf$
affects the net probability by $<$10\%, since the singly-imaged
case is dominated by clusters.

\reffig{mult1} shows the multiply-imaged magnification probability,
for different assumptions about the dwarf/galaxy transition and
different single-detectable-image criteria (SDIC).  In general,
nearly all of the probability comes from galaxies.  Clusters
contribute $<\!10^{-5}$ of the probability if we consider all
image configurations, and $<\!10^{-8}$ if we consider those
with only a single detectable image; so we confirm the result from
\citet{CHS} that clusters are essentially negligible for the
multiply-imaged magnification probability.  NFW dwarfs halos, if
present in the model, likewise have a negligible contribution to the
probability (no more than $\sim\!10^{-7}$ even in the model with
$\log\Mdwarf = 12.5$).  The main effect of having dwarfs in the model
is to reduce the number of halos that are isothermal galaxies, and
hence to reduce the net multiply-imaged magnification probability.
The reduction can be substantial when the SDICs are important.

Perhaps the most interesting result from \reffig{mult1} is the effects
of the SDICs.  Consider the probability of a magnification $\mu
> 10$, and suppose we have no knowledge of the presence or absence of
additional images.  Then we must use the multiply-imaged probability
with no SDICs, and add the singly-imaged probability, which
yields total probabilities of $(2.1, 1.4, 1.0) \times 10^{-4}$ for
models (a), (b), and (c) respectively.  However, suppose like
\citet{richards} we can rule out the presence of extra images down to
either $\Dt=0\farcs1$ and $f=0.1$, or $\Dt=0\farcs3$ and $f=0.01$.  We
should then use the lowest of the multiply-imaged curves in
\reffig{mult1}, and again add the singly-imaged probability.  A useful
way to quantify the results is to give the fraction of systems
magnified by $\mu>10$ that do not have extra images detectable by
\citet{richards}.  This fraction is given by the {\em ratio} of the
probability with SDICs to the probability without, or
differences between the curves in \reffig{mult1}; the results are
given in \reftab{frac}.  For example, in the no-dwarf model, about
24\% of highly magnified systems are multiply-imaged such that the
extra images are undetectable, and another 5\% are singly-imaged.  In
models with dwarfs, the fraction that are multiply-imaged without
detectable extra images is much lower, so in total only 9--15\% of
highly magnified systems lack detectable extra images, and many of
those are true singly-imaged systems lensed by clusters.

\begin{deluxetable}{lrrr}
\tablecaption{}
\tablehead{
  \colhead{Model} &
  \colhead{$F_{\rm sing}$} &
  \colhead{$F_{\rm mult}$} &
  \colhead{$F_{\rm tot}$}
}
\startdata
  no dwarfs          & 0.049 & 0.241 & 0.290 \\
  $\log\Mdwarf=11.0$ & 0.059 & 0.090 & 0.149 \\
  $\log\Mdwarf=12.0$ & 0.080 & 0.014 & 0.094 \\
  $\log\Mdwarf=12.5$ & 0.107 & 0.008 & 0.115 \\
\enddata
\tablecomments{
Fraction of lens systems magnified by $\mu>10$ that lack extra
images detectable by \citet{richards}.  Columns 2-3 give the
fractions that are singly-imaged and multiply-imaged, respectively,
and Column 4 gives the total.  Here we give results for a model
with the dwarf/galaxy transition at $\log\Mdwarf = 11.0$, in
addition to the three models discussed in the text.
}\label{tab:frac}
\end{deluxetable}

\begin{figure*}
\centerline{\epsfxsize=6.2in \epsfbox{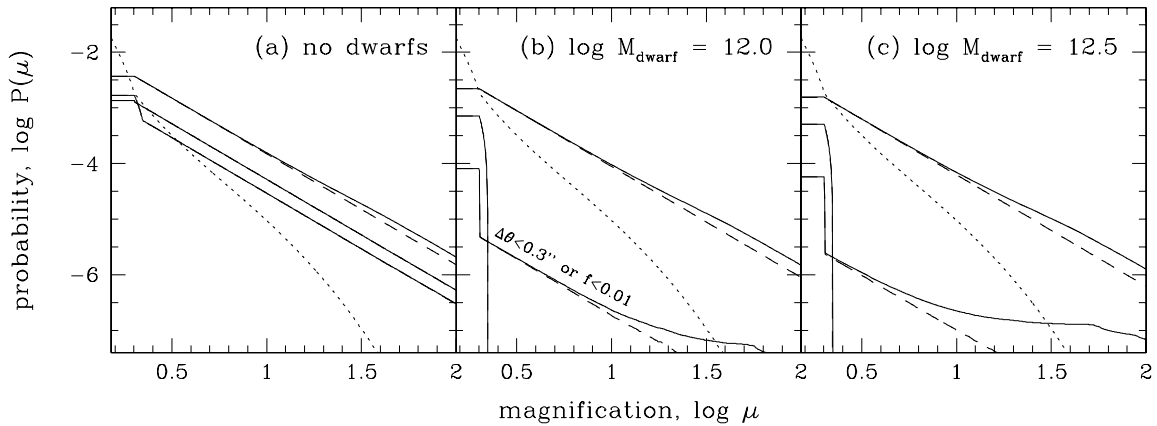}}
\caption{
Similar to \reffig{mult1}, but for a model in which all halos are
spherical and there is no shear.
}\label{fig:mult-sph}
\end{figure*}

\subsection{Implications for the SDSS quasars}
\label{sec:SDSS}

So far we have only considered {\em a priori} lensing probabilities,
i.e., the bare optical depth for producing certain magnifications.
However, to compute the probability of finding a certain magnification
in a real, flux-limited survey we would also have to fold in magnification
bias to obtain {\em a posteriori} probabilities.  Because the SDSS can
probe only the steep, bright end of the quasar luminosity function at
$z \approx 6$, magnification bias may be quite strong, but it is is
extremely sensitive to the poorly-known LF shape.  (In fact, the
problem can be turned around so that lensing [or lack thereof] in the
$z \approx 6$ quasar sample yields constrains on the LF slope; see
\citealt{CHS} and \citealt{richards}.)  Rather than making detailed
but highly LF-dependent predictions of the {\em a posteriori}
probabilities, we turn attention to probability {\em ratios} and
consider the following question: What is the ratio of the probability
for $\mu>10$ with observational SDICs, to the probability for
$\mu>10$ without SDICs?  This is equivalent to the question: In
a toy model with magnification bias so strong that {\em all} $z
\approx 6$ quasars are magnified by a factor $\mu>10$, what is the
probability that one of the quasars would have no additional images
detectable by \citet{richards}?

In general, probability {\em ratios} should be the same for
{\em a priori} and {\em a posteriori} probabilities, so
\reftab{frac} is exactly what we need.  To summarize, if we are
extremely optimistic about having strong magnification bias and
about all low-mass halos having steep isothermal profiles, we can
imagine that the probability of a $z \approx 6$ quasar being
magnified by $\mu>10$ without having additional images detectable
by HST might be as high as 29\%.  If this is the case, we cannot
rule out the possibility that one of the four $z \approx 6$ quasars
observed by \citet{richards} might actually be significantly
magnified.  However, even then the probability that {\em all four}
are amplified would be $P = (0.29)^4 = 0.007$.  The actual
probability is almost certainly much lower, because magnification
bias is probably not as strong as in the toy example, and because
assuming that all low-mass halos have isothermal profiles probably
overestimates the lensing optical depth.  In other words, we can
rule out at more than 99.3\% confidence the hypothesis that all
four $z \approx 6$ quasars are amplified by more than a factor of
10 --- provided we can equate {\em a priori} and {\em a posteriori}
probability ratios.

The only possible problem with equating the probability ratios is
if the SDSS is somehow biased against high-redshift lensed quasars.
Suppose, for the sake of argument, that there were a total of
$4/0.29 = 14$ lensed and highly magnified $z \approx 6$ quasars
in the SDSS, 10 of which have not been identified.  Then having
four quasars that are magnified but lack images detectable by
\citet{richards} would actually be consistent with the no-dwarf
model.  In our models with dwarfs, the total number of lensed
quasars needed would be $\sim$25--45 (see \reftab{frac}).  The key
question is whether so many ``missing'' lensed high-redshift
quasars could exist.  The SDSS $z>5.8$ quasar sample is selected
on the basis of colors alone; the sample is basically $i$-dropouts,
with some additional color criteria to reduce contaminants
\citep{fetal00,fetal01,fetal03}.  There is no requirement that
the objects appear point-like in SDSS images, and hence no bias
against multiply-imaged systems.  The only remaining problem is
if light from a lens galaxy could change the composite colors of
a lensed quasar system enough that the system would not be selected
as an $i$-dropout.  Lens galaxies associated with $z \approx 6$
lensed quasars would be expected to lie at redshifts
$1 \lesssim z_l \lesssim 2$ \citep[see Fig.~1 of][]{CHS}, and so
would probably be too faint to significantly change the colors
\citep[also see][]{WL02}.  While it would be interesting to quantify
this effect more carefully,\footnote{An analysis that begins with
the empirical correlation between image separation and lens galaxy
luminosity \citep[see the Appendix of][]{rusin-evol} would be
reasonably straightforward.} it seems unlikely that the SDSS
high-redshift quasar sample is highly biased against multiply-imaged
systems.

\subsection{Corollaries and Systematics}
\label{sec:systematics}

We have obtained our main result, namely answering the question of
whether it is likely that the $z \approx 6$ quasars are highly
magnified, but there are several corollaries worth mentioning.  First,
so far in this section we have allowed the galaxies to have
ellipticity and shear.  (We have made certain assumptions about the
ellipticity and shear for dwarf and cluster halos, but they are not so
important; see \refsec{popmod}.)  It is interesting to repeat the
analysis with all halos assumed to be spherical, to see how much
ellipticity and shear affect the results.  \reffig{mult-sph} shows the
results.  In the no-dwarf model, neglecting ellipticity and shear
reduces the probability of multiple imaging with $\mu>10$ by 30--50\%.
The same holds in models with dwarfs for the total multiple imaging
probability (with no selection effects).  However, in models with
dwarfs where the SDICs apply, the reduction is at least a factor
of 5 and often much larger.  This result can be understood as follows.
Most of the multiply-imaged magnification probability comes from the
isothermal halos called ``galaxies'' in our model.  If there are no
dwarfs, then most of the probability comes from low-mass galaxies that
produce image separations too small to be resolved; so the flux ratio
SDIC is unimportant, and the probability is not dramatically
sensitive to ellipticity (see the upper curves in
\reffig{iso-dist2e}).  By contrast, if low-mass halos are dwarfs
(which have negligible cross sections), then much of the probability
must come from galaxies that produce image separations larger than the
resolution.  In this case the flux ratio SDIC plays an important
role; and in \reffig{iso-dist2e} we saw that this dramatically reduces
the cross section when the ellipticity is zero.  The bottom line is
that ellipticity and shear are a factor of $\sim$2 effect in no-dwarf
models, but can be an order of magnitude effect in models with dwarfs.

The second point is that there are several additional systematic
uncertainties that might be relevant.  We have considered three
effects that have the most impact on strong lens statistics
\citep[see][]{kuhlen}: (i) changing the halo mass function from
that of \citet{ST} to that of \citet{jenkins}; (ii) changing the
scatter in the $M$-$c$ correlation for NFW halos from 0.14~dex
to 0.07~dex or 0.21~dex; and (iii) varying the location of the
cluster/galaxy transition by $\Delta(\log\Mclus) = \pm0.25$.  We
find that changing the mass function has the strongest effect, and
even that is only a $\sim$10\% change in the net magnification
probability.  In other words, our results appear to be robust to
effects other than the question of whether low-mass halos are
isothermal galaxies or NFW dwarfs.

\begin{figure}
\centerline{\epsfxsize=3.1in \epsfbox{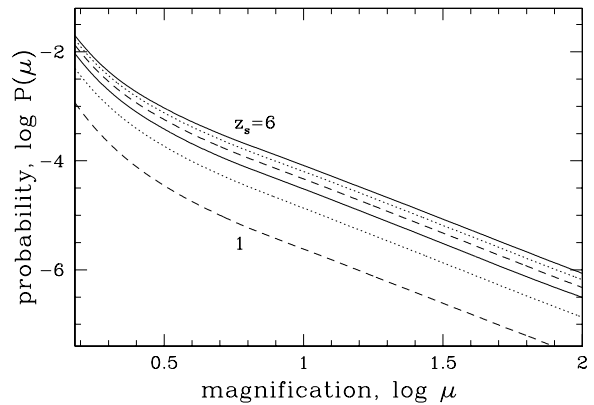}}
\caption{
Total magnification probability distributions, including both the
singly-imaged and multiply-imaged contributions, as a function of
source redshift $z_s=6,5,4,3,2,1$ from top to bottom.  Here we
use the no-dwarf model to obtain the maximum possible probability.
}\label{fig:zsrc}
\end{figure}

Finally, our discussion has been geared toward sources at $z \approx
6$, but for completeness in \reffig{zsrc} we show the net {\em a
priori} magnification probability as a function of source redshift.
We use the no-dwarf model as a way to obtain an upper bound on the
probability; for the multiply-imaged case we use the criteria
$\Dt<0\farcs3$ or $f<0.01$ for detecting only a single image,
although from \reffig{mult1}a this choice is not so important.
Reducing the redshift naturally reduces the probability, especially
for $z_s \lesssim 3$.  It affects the whole distribution in the same
way, so our general conclusions appear not to be highly sensitive to
the source redshift.  Pushing quasar and galaxy samples beyond $z
\approx 6$ will not significantly increase the probability for large
lensing magnifications.

\section{Conclusions}

The problem of lensing magnification without multiple detectable
images has a rich phenomenology.  First, there is the case of true
singly-imaged systems.  Isothermal halos are not very efficient
at producing highly magnified single-image systems, but NFW halos
are.  Consequently, high singly-imaged magnifications are possible
in principle, and they are mainly associated with massive
($\gtrsim\!10^{13.5}\,M_\odot$) halos corresponding to clusters
of galaxies.  An important implication for observations is that,
if there is no evidence for a cluster along the line of sight to
a distant quasar, then it is unlikely that there is strong
singly-imaged magnification.

The second case is when there are multiple images but the extra
images are not detectable, either because the image splitting is
too small to be resolved, or because the extra images are too faint.
NFW halos are inefficient at producing multiple images, and when
they do they rarely produce extreme flux ratios; therefore,
clusters contribute negligibly to the multiply-imaged magnification
probability.  Instead, this case is dominated by galaxies and
lower-mass systems ($\lesssim\!10^{13}\,M_\odot$).  The probability
is very sensitive to the inner density profile of these halos.
If all low-mass halos have steep isothermal profiles, then the
probability is dominated by lens systems with image separations
too small to be resolved by HST.  However, if low-mass dwarf
halos have NFW profiles (and hence small cross sections), then
the overall probability is dominated by lens systems where the
extra images are faint.

Our central quantitative result is that 9--29\% of all lens
systems with magnifications $\mu>10$ lack additional detectable
images.  In a toy model where magnification bias is so strong
that most or all $z \approx 6$ quasars are lensed, then we cannot
rule out the hypothesis that one of the four quasars observed by
\citet{richards} is magnified despite lacking extra images.
However, even in such an extreme model, the probability that
{\em all four} are magnified by a factor of 10 would still be no
more than 0.7\%, and is probably much lower.  The only way to
evade this argument is if the SDSS high-redshift quasar sample is
somehow biased against quasar lens systems with multiple bright
images.  That seems unlikely, although a detailed analysis of
whether light from a lens galaxy could cause the $i$-dropout
selection technique to miss a lens system is needed to answer
this question definitively.

Incidentally, we can comment on the two different criteria used
by \citet{richards} to search for companion images to the
$z \approx 6$ quasars.  \citeauthor{richards} were able to rule
out extra images down to a flux ratio $f=0.01$ at image splittings
greater than $0\farcs3$, or down to a less stringent flux ratio
$f=0.1$ for smaller image splittings down to $0\farcs1$.  We find
that for the no-dwarf model, the two criteria give similar
probability results.  By contrast, for models with dwarfs, the
criteria with more stringent flux bounds provide stronger
constraints on the lensing probabilities.  In other words, for
the purpose of determining whether distant quasars are magnified
by lensing, it is more valuable to aim for more dynamic range than
to push for resolution much better than $\sim\!0\farcs3$.

To summarize, if the SDSS high-redshift quasar sample is not
highly biased against multiply-imaged quasars, then it is quite
improbable that all four quasars observed by \citet{richards} are
highly magnified.  In that case, the quasars can be taken as good
evidence for the presence of billion-$M_\odot$ black holes in the
young universe.  Explaining such black holes is a challenge for
black hole growth models, whose solution may involve a need for
super-Eddington accretion \citep{zh04}.  In other words, in the
case of high-redshift quasars, a lensing ``null result'' actually
makes the objects even more interesting.

\acknowledgements
We thank Gordon Richards for helpful discussions about the SDSS
high-redshift quasar sample.
CRK is supported by NASA through Hubble Fellowship grant
HST-HF-01141.01-A from the Space Telescope Science Institute, which
is operated by the Association of Universities for Research in
Astronomy, Inc., under NASA contract NAS5-26555.
MQK is supported by NSF grant AST-0205738.
ZH is supported by NSF grants AST-0307200 and AST-0307291.

\appendix

\section{Maximum Singly-Imaged Magnification for Isothermal Lenses}

With isothermal lenses there are three simple cases in which the
maximum singly-imaged magnification can be obtained analytically;
the bounds were are shown in \reffig{iso-max} and are derived here.
First, for a simple isothermal sphere the magnification of a
singly-imaged source ($u>\Rein$) is $\mu = 1 + \Rein/u$, so the
maximum magnification for singly-imaged sources is $\mumax=2$.
In fact, in this case the full singly-imaged magnification cross
section can be derived analytically,
\begin{equation}
  A(\mu) = \pi \Rein^2 \left[ \frac{1}{(\mu-1)^2} - 1 \right] .
  \qquad(1<\mu<2)
\end{equation}

Next consider an isothermal sphere with external shear.  The
magnification as a function of position is
\begin{equation}
  \mu^{-1} = 1 - \gamma^2 - \frac{\Rein}{R}\left(1+\gamma\cos2\theta\right)\,.
\end{equation}
(We are now working in a coordinate system aligned with the shear,
so $\theta_\gamma=0$ in these coordinates.)  The radial caustic is a
circle with radius $\Rein$.  It maps to a curve in the image plane
called the 1-2 transition locus, which marks the transition from
single images to images that are part of a two-image system, and
which can be written in polar coordinates as \citep{finch}
\begin{equation}
  R_{1-2}(\theta) = 2\Rein \left(\frac{1+\gamma\cos2\theta}
    {1+2\gamma\cos2\theta+\gamma^2}\right) .
\end{equation}
For $\gamma<1/3$ the radial caustic completely encloses the
tangential caustic, so all sources outside the radial caustic and
all images outside the 1-2 transition locus are singly-imaged.
In this case, the maximum singly-imaged magnification occurs at
$\theta=0$ on the 1-2 transition locus,
\begin{equation}
  \mumax = \frac{2}{(1-3\gamma)(1+\gamma)}\ .
  \qquad(\gamma<1/3)
\end{equation}
For $\gamma>1/3$ the tangential caustic pierces the radial caustic
to form a naked cusp \citep[e.g.,][]{SEF,finch}.  Sources just
outside a naked cusp are singly-imaged but can have arbitrarily
large magnifications, so $\mumax \to \infty$ in this case.

Finally consider an isothermal ellipsoid.  The magnification as a
function of position is
\begin{equation}
  \mu^{-1} = 1 -
    \frac{b}{R}\,\left[\frac{2}{(1+q^2)-(1-q^2)\cos2\theta}\right]^{1/2} .
\end{equation}
We cannot compute the full 1-2 transition locus, but we can compute
where it intersects the $x$ and $y$ axes:
\begin{eqnarray}
  R_{1-2}(0) &=&
    \frac{2b}{\sqrt{1-q^2}} \tan^{-1}\left[\frac{\sqrt{1-q^2}}{q}\right] \\
  R_{1-2}(\pi/2) &=&
    \frac{2b}{\sqrt{1-q^2}} \mbox{tanh}^{-1}\left[\sqrt{1-q^2}\right]
\end{eqnarray}
The critical value of the axis ratio at which the tangential caustic
pierces the radial caustic to form a naked cusp is found by solving
\begin{equation}
	2 \tan^{-1}\left[\frac{\sqrt{1-q^2}}{q}\right] - \frac{\sqrt{1-q^2}}{q} = 0\,,
\end{equation}
whose solution is $q=0.394$, or ellipticity $e=0.606$.  For $q<0.394$,
the presence of a naked cusp again causes $\mumax \to \infty$.  For
$q>0.394$, the maximum singly-imaged magnification again occurs at
$\theta=0$ on the 1-2 transition locus and has the value
\begin{equation}
  \mumax = \frac{q\,R_{1-2}(0)}{q\,R_{1-2}(0)-b}\ .
  \qquad(q>0.394)
\end{equation}

\end{document}